\documentclass[apjl]{emulateapj}
\usepackage{apjfonts}

\shorttitle{Carina Tidal Tails}
\shortauthors{Mu\~noz et al.}

\begin{document}

\title{Exploring Halo Substructure with Giant Stars XI:
The Tidal Tails of the Carina Dwarf Spheroidal
and the Discovery of Magellanic Cloud Stars in the Carina Foreground}

\author{
Ricardo R. Mu\~noz\altaffilmark{1},
Steven R. Majewski\altaffilmark{1}, 
Simone Zaggia\altaffilmark{2},
William E. Kunkel\altaffilmark{3}, 
Peter M. Frinchaboy\altaffilmark{1},
David L. Nidever\altaffilmark{1},
Denija Crnojevic\altaffilmark{2},
Richard J. Patterson\altaffilmark{1,4}, 
Jeffrey D. Crane\altaffilmark{1,4,6},
Kathryn V. Johnston\altaffilmark{5},
Sangmo Tony Sohn\altaffilmark{1,7},
Rebecca Bernstein\altaffilmark{8}, \&
Stephen Shectman\altaffilmark{6} 
}

\altaffiltext{1}{Dept. of Astronomy, University of Virginia,
Charlottesville, VA 22903-0818 (rrm8f, srm4n, pmf8b, ricky, dln5q@virginia.edu)}

\altaffiltext{2}{Osservatorio Astronomico di Trieste, Via G. B. Tiepolo 11,
I-34131 Trieste, Italy (zaggia@oats.inaf.it)}

\altaffiltext{3}{Las Campanas Observatory, Casilla 601, La Serena, Chile
(kunkel@jeito.lco.cl)}

\altaffiltext{4}{Visiting Astronomer, Cerro Tololo InterAmerican Observatory, National
Optical Astronomy Observatories}

\altaffiltext{5}{Department of Astronomy, Wesleyan University, Middletown, CT
(kvj@urania.astro.wesleyan.edu)}

\altaffiltext{6}{Carnegie Observatories, 813 Santa Barbara Street,
Pasadena, CA 91101 (crane@ociw.edu,shectman@ociw.edu)}

\altaffiltext{7}{Korea Astronomy and Space Science Institute, 61-1 Hwaam-Dong,
Yuseong-Gu, Daejeon 305-348 Korea (tonysohn@kasi.re.kr)}

\altaffiltext{8}{Dept. of Astronomy, University of Michigan,
Ann Arbor, MI 48109 (rabernst@umich.edu)}

\begin{abstract}
 
A new large-area Washington $M,T_2+DDO51$ filter survey of more 
than 10 deg$^2$ around the Carina dSph galaxy reveals a 
{\it spectroscopically confirmed} power law
radial density ``break" population of Carina giant stars
extending several degrees beyond the central King profile.
Magellan telescope MIKE spectroscopy establishes
the existence of Carina stars to at least
4.5 times its central King limiting radius, $r_{lim}$ and primarily
along Carina's major axis.
To keep these stars bound to the dSph would require a 
global Carina mass-to-light ratio of $M/L \ge 6,300$ (M/L)$_{\sun}$.
The MIKE velocities, supplemented with 
$\sim950$ additional Carina field velocities 
from archived VLT+GIRAFFE spectra with $r \lesssim  r_{lim}$,
demonstrate a nearly constant Carina velocity dispersion 
($\sigma_{\rm v}$) to just beyond $r = r_{lim}$, 
and both a rising $\sigma_{\rm v}$ and a velocity shear at still larger radii.
Together, the 
observational evidence
suggests that the discovered extended Carina population
represents tidal debris from the dSph.
Of 65 giant candidates at large angular radii from the Carina center
for which MIKE spectra have been obtained 94\% are 
associated either with Carina or a second, newly discovered diffuse, but
strongly radial velocity-coherent ($\sigma_{\rm v}$=9.8 km s$^{-1}$), 
foreground halo system.
The fifteen stars in this second, retrograde velocity population have
(1) a mean metallicity $\sim1$ dex higher than that of Carina, and 
(2) colors and magnitudes 
consistent with the red clump of the Large Magellanic Cloud (LMC). 
Additional spectroscopy of 
giant star candidates in fields linking Carina and the LMC show a smooth
velocity gradient between the LMC and the retrograde Carina moving group.
We conclude that we have found Magellanic 
stars almost twice as far (22$^{\circ}$) from the LMC center than
previously known.  
 
\end{abstract}

\keywords{Carina Dwarf -- galaxies: Local Group -- kinematics and dynamics 
-- Magellanic Clouds --cosmology : dark matter }

\section{Introduction}

Whether the Milky Way dwarf spheroidal (dSph) satellite galaxies are undergoing tidal disruption
remains a controversial question.  Such tidal
disruption would naturally lead to extended populations of stars that have been stripped 
from the satellite core.  That {\it most} of the Milky Way (MW) dwarf spheroidals exhibit 
radial density profiles with extended components was suggested by the 
large area photographic survey of most of the Galactic dSph satellites by
\citet[][hereafter {IH95}]{IH95}.  A number of studies have addressed the question of the reality of 
extended structural components around individual dSph examples ---
among them the Carina dSph, for which the issue has prompted 
lively debate  (\citealt{kuhn96}, \citealt[][hereafter {Paper II}]{Paper II}, \citealt{mor01}, 
\citealt{Walcher2003}, \citealt{monelli04}).  
In a recent review of past and new work on the Carina system, 
\citet[][hereafter {Paper VI}]{Paper VI} attempted to resolve the previous, apparently 
discordant results 
regarding the photometric detection of an extended Carina structural component.  
Paper VI showed that of all previous photometric surveys of Carina, 
that of \citeauthor*{Paper II}
--- which makes use of the $DDO51+$Washington $M,T_2$ filter technique to identify giant
stars (\citealt[][hereafter {Paper I}]{Paper I}) at the distance of the 
Carina system --- achieves the highest,
and therefore most reliable, 
signal-to-background contrast in the diffuse outer parts of the Carina system. 
Moreover an extended, power-law component detected around Carina in \citeauthor*{Paper II}
is supported by spectroscopic
confirmation of Carina giant candidates to 1.4 times the nominal
limiting radius ($r_{lim}$) 
of the central-fitted King profile in Paper VI.  Our previous work has therefore established
the likely reality of the ``King + power law" density profile for the Carina dSph.

In this paper (\S3) we take advantage of a similar, but deeper and much
wider area, photometric database of Carina than that presented in \citeauthor*{Paper II} and,
in addition,
contribute higher-quality radial velocities (RVs) of Carina stars from 
echelle spectroscopy of giant star candidates to more than three times the 
angular separation from the Carina center than we
explored in \citeauthor*{Paper VI}. 
Carina-associated stars are now established to 
$4.5 r_{lim}$ from the Carina core, leaving no doubt as to 
the reality of an extended component to the Carina system and imposing extreme 
limits on the mass of Carina if these stars are bound to the dSph (\S4.6).  
To further improve the kinematical mapping of the Carina system at smaller radii, 
we also take advantage of archived, publicly available VLT+GIRAFFE 
spectra of more than
1000 stars near the center of Carina, which 
contribute more than 300 additional RVs of Carina-related stars 
within the King limiting radius.
The resulting velocity dispersion profile of the Carina system is the most extensive yet 
determined for any dSph galaxy, yet shows a continuation to large radii of the same more 
or less flat trend found
(to smaller radii) in other Galactic dSphs (\citealt{Mu05};
\citealt{Westfall2006}; \citealt{Walker2005}; \citealt{Sohn2006}) .

The present photometric and spectroscopic database of stars in the direction of 
Carina has yielded the additional discovery of a second apparently coherent stellar
population in the foreground of the dSph (\S6).  This other Milky Way substructure is 
as dynamically cold as the Carina system itself and, ironically, represents the primary 
source of contamination within our previous (\citeauthor*{Paper II}) and present 
photometric samples of Carina stars outside $r_{lim}$.
The fifteen stars in our MIKE sample that are part of
 this other substructure share a number of properties (color-magnitude
diagram position, metallicity, and velocity-angular separation trend) 
with stars of the Large Magellanic Cloud (LMC), but stretch
some 22$^{\circ}$ from the LMC center.  As with Carina, these widely separated stars
place new, very large lower limits on the LMC mass and tidal radius if the stars are bound to
their parent satellite.

\section{New Photometric Survey} 

\subsection{Imaging Data }

Through spectroscopic follow-up of stars in the \citeauthor*{Paper II} Carina database,
\citeauthor*{Paper VI} demonstrated the efficacy of Washington $M,T_2+DDO51$ 
photometry to produce high quality candidate lists of giant stars from the Carina system
to large separations from the core, and dispelled concerns 
(\citealt{mor01}; \citealt{Mayer2002}; \citealt{Walcher2003})
that there may have been problems with the original methods or 
findings of \citeauthor*{Paper II}.
Nevertheless, a deeper, more uniform, and larger area
$M,T_2+DDO51$ survey of the Carina system was desired: \citeauthor*{Paper VI} (see \S 2.4 
of that paper) showed 
how the results of a survey with better photometry would improve the Carina giant candidate 
selection, whereas surveying to a larger angular radius would give greater insight into the
extent and 
character of this outer Carina population.
 
Thus, new Carina photometry 
over a 10.74 deg$^2$ area (9.3 times more area than covered in \citeauthor*{Paper II} ---
the area outlined below in Fig.\ 1) centered on the Carina dSph
was obtained with the Mosaic wide-field imaging camera on the Blanco telescope  
on UT 2000 Feb 24-27 under photometric conditions.
DAOPHOT II/ALLSTAR (\citealt{Stetson1987}) PSF-fitting photometry 
was derived for stars in each of the individual Mosaic pointings, 
producing magnitudes with
median errors of $(\sigma_M,\sigma_{T_2},\sigma_{DDO51}) =$ 
(0.018, 0.020, 0.015) at $M=20.8$, which is approximately 3.4 mag below the Carina
red giant branch (RGB) tip (Fig.\ 2).  The photometry of this new survey is about 
2 times more precise at that magnitude
than the Carina data presented in \citeauthor*{Paper II}.  Instrumental 
magnitudes were calibrated into the standard system via
multiple observations of Washington$+DDO51$ standards in \citet{geisler96}.
Each star in our catalog has been corrected for reddening based on 
its Galactic coordinates by using the reddening map constructed by
\citet{Schlegel98}. We found an $E(B-V)$ range of 0.033--0.102 in our Carina fields.

\subsection{Carina dSph Candidate Selection and Density Profile}

As in \citeauthor*{Paper II}, the dereddened $(M-T_2,M-DDO51)$ two-color diagram (2CD)
and the $(M-T_2,M)$ color-magnitude diagram (CMD)
are used together to select stars most likely to be Carina giant stars.  Figure 3
illustrates the selection criteria we used to identify Carina RGB stars:
``Carina giant star candidates" are expected to fall primarily
within the regions bounded by the solid lines 
in each of the diagrams in Figure 3. 
Because we want to create the most reliable maps of Carina density possible and because
we choose to reserve the valuable Magellan echelle spectroscopic follow-up (\S3) observing 
for the very best photometrically-selected candidates,
our initial selection criteria were deliberately conservative.
For example, 
we did {\it not} employ the proposed wider
CMD selection criteria discussed in \S3.2 of \citeauthor*{Paper VI}, but maintained 
the more restrictive limits used in \citeauthor*{Paper II}. 
In addition, our 2CD giant selection boundary is set far from the dwarf star locus to minimize
photometric contamination (Fig. 3b) of the giant sample.
However, the extensive, archived VLT+GIRAFFE spectroscopic data set 
for the Carina field, obtained for dSph candidates selected independently
of our photometry and methodology, allows us in the velocity analysis described later (\S4)
to search for additional Carina stars with measured RVs that, while being excluded 
from our conservatively made ``best" candidate lists, still occupy "RGB-like"
regions of the CMD and 2CD (\S 4.1.3).

Across our survey area, the photometric sample is expected to be complete to $M=20.8$,
so we analyze the spatial distribution of giant candidates to this magnitude limit.
In addition, 
because our spectroscopic survey is almost complete outside the Carina $r_{lim}$
to $T_2 = 18.4$, we also analyze the spatial distribution of 
giants using this magnitude limit.
Figure 4 presents the radial density profile derived for the Carina dSph for these
two adopted magnitude limits.  To create this profile, stars have been binned into
elliptically-shaped annuli matching the Carina center, ellipticity and position angle found by 
\citeauthor*{IH95}.

As discussed at length in \citeauthor*{Paper VI} and \citet{Westfall2006}, 
proper assessment of the background level (i.e. density of false positive detections) 
is critical to deriving dSph radial density profiles.  Here we adopt 
two strategies for
assessing this backgound level.  In the case of the $T_2$-limited density profile, we
can very accurately estimate the background directly from the results of our spectroscopic
survey (\S 4), which is 90\% complete for stars beyond the \citeauthor*{IH95} Carina $r_{lim}$.
This spectroscopically-verified, true background\footnote{This is the background level 
scaled to a 100\% spectroscopic completeness level.} 
of 2.3 deg$^{-2}$
within our ``Carina giant candidate" star sample 
is subtracted from the observed density distribution of 
$T_2 \le 18.4$ Carina dSph giant candidates across the entire survey to reveal the density profile
shown in Figure 4a.  Because of our near spectroscopic completeness for these stars, the
density profile shown in Figure 4a beyond the \citeauthor*{IH95} $r_{lim}$  
virtually reflects the exact distribution of all RV-verified members there.

For our $M \le 20.8$ sample, the background is estimated 
using the ``CMD-shifting" method used in \citeauthor*{Paper II} to estimate a background rate,
with the one important difference that with our new survey 
here we are able to make exclusive use of the vast area outside of
the Carina King limiting radius to significantly reduce potential contribution of 
any Carina stars not lying on the Carina RGB (e.g., asymptotic and 
post-asymptotic giant branch stars) that may have 
artificially inflated the estimated backgrounds in the \citeauthor*{Paper II} execution of this method. 
On the other hand, we acknowledge that this method may also underestimate 
the contribution of the newly discovered halo substructure discussed in \S6, since
it has a similar CMD position as the Carina dSph.
To correct for this, we add back into our background estimate 
the fractional contribution of stars from this substructure among
the spectroscopic sample of $>r_{lim}$ stars chosen as Carina giant candidates.
This yields a conservative\footnote{The procedure just described {\it ignores}
the fact that much of the \S 6 substructure is actually {\it outside} our CMD selection
criterion, to make the most generous estimate of the background.}
 background level of 10.3 deg$^{-2}$, which is then subtracted from the 
observed density distribution of $M \le 20.8$ Carina RGB candidates. 

The two samples with the two methods of background calculation produce remarkably
consistent radial profiles (Fig. 4).
In both cases, the central part of the density profile is 
well described by the normalized \citeauthor*{IH95} 
King profile (shown by the solid line) for Carina, which
is characterized by $r_{lim} = 28.8$ arcmin and a core
radius of 8.8 arcmin.
Both radial distributions also show a prominent, second ``break population" roughly following
a power-law decline 
to the limits of our present survey.  
The dashed lines correspond to power-law indices of -1.5, -2 and -2.5 respectively.  
A -2 index power law appears to yield a reasonable match to the density fall-off of the
break population, although power laws with indices of -1.5 or -2.5 cannot be discounted; 
in general, the power law here is steeper than found in \citeauthor*{Paper II}, owing 
to a slightly higher 
background derived in this study (see
also a discussion of this steeper slope in \citeauthor*{Paper VI}).
Nevertheless, in this completely new photometric survey with substantial spectroscopic
follow-up we have independently borne out the general conclusion of \citeauthor*{Paper II}
and Paper VI that Carina exhibits a prominent, extended, power-law break population.

As discussed in \citeauthor*{Paper II}, the density profile exhibited in Figure 4 mimics that
of model disrupting dSph galaxies (see, e.g., Fig.\ 15 of \citealt{JSH99}).  
With our updated version of the Carina density profile, we
can revisit the implied fractional mass loss rate according to the method of
\citet{JSH99} under the assumption that the power law population 
represents unbound tidal debris. We derive a fractional mass loss rate for Carina of 
$(df/dt)_{1}$=0.075 Gyr$^{-1}$, but we must note that this method is technically
derived for break populations following a -1 power law, and even in that case it only 
yields estimates good to within a factor of two.
Perhaps a better estimate of the fractional mass loss rate comes from the $(df/dt)_2$ method
of \citet{JSH99} using the corrections given by 
\citet{JCG02};
this method yields un upper limit for the Carina mass loss rate of 
$(df/dt)_{2} < 0.24$ Gyr$^{-1}$.
In a subsequent paper (Mu\~noz et al., in preparation, hereafter M06) we use this newly derived
density profile as well as the velocity dispersion profile derived in \S 4.3 to model 
the mass loss history using N-body simulations specific
to the Carina dSph, and derive likely mass loss rates generally between these two
estimates.

\section{Spectroscopic Data}

\subsection{Spectroscopy with MIKE}

The Carina power-law population has been of particular, though not exclusive, 
interest during 
our follow-up spectroscopic observations.
\citeauthor*{Paper VI} presented radial velocity observations obtained with the Blanco telescope + Hydra
multifiber system. Only some of these observations were of sufficient resolution to contribute
reliable information on the internal dynamics of (rather than simply stellar membership in) 
the Carina dSph.
Thus, on UT 2004 Jan 27-28 and Dec 29-30 good spectra of 
a total of 77 Carina giant star candidates 
selected from the new 
photometric survey were obtained using the Magellan Inamori Kyocera Echelle (MIKE)
spectrograph on the Clay 6.5-m telescope at Las Campanas; this instrument delivers $R\sim19000$
resolution spectra over the red echelle orders we used for this work. 
Of the 77 stars targeted,
65 were selected to be giant candidates by the 
giant selection shown by the solid lines in Figure 3 
and 12 were selected using an expanded 2CD 
selection (shown by the dotted lines in the same figure). This wider 2CD giant
selection corresponds to that derived in \citeauthor*{Paper I}.
More than half (40) of the 77 stars observed with MIKE
lie outside the nominal Carina $r_{lim}$
as determined by \citeauthor*{IH95}; the rest are scattered throughout the 
region inside the King limiting radius, but
primarily at larger radii where few previous Carina spectra have been obtained.

Radial velocities (RVs) have been derived via cross-correlation of the MIKE spectra
against a ``universal template" containing sets of stellar atmospheric absorption lines that 
typically give the strongest correlations to the spectra of late type stars; apart from these
lines, the bulk of the spectra and templates are masked out because these wavelengths
contribute more noise than signal to the cross-correlation spectrum.  
Prior to cross-correlation, the spectra are
also Fourier-filtered to remove irrelevant low frequency features as well as features with
higher frequency than the intrinsic resolution of the spectrograph.  A fuller discussion
of this cross-correlation technique is given in \citet{Majewski2004a}; but we have found that
the procedure works just as well, or even better, for $R=19,000$ spectra than for the 
moderate resolution spectra cross-correlated in that paper.
We observed to high $S/N$ a number of 
K giant velocity standard stars that we used to measure small systematic offsets imposed on 
the derived RVs that are particular to the nature of the adopted artificial template.
Our cross-correlations here were conducted over the echelle order (spanning 8468-8693 \AA\ )
that contains the calcium infrared triplet
and over a dozen other useful lines in stars as metal poor as Carina ([Fe/H] $\sim -2$).
Tests with other, nearby orders yield similar RV results but of lower reliability, so
the values given here are based solely on the calcium triplet order, where the typical
$S/N$ of the stellar continua were 7-12 per pixel.  
This particular echelle order also contains
ample numbers of telluric absorption features with strengths great enough
to yield useful velocities.
Since the stars were observed
with a 0.9 arcsec slit whereas the seeing often was as good as 0.7 arcsec, significant
fractional errors in derived RVs may arise from slit centering errors. To measure
the velocity shifts that result from this effect, we independently cross-correlate the 
telluric absorption features in each order against those in a set of 
observed RV standards as well as in
dusk spectra (see discussion in \citealt{Sohn2006}). 
These RV standards were typically exposed by smoothly passing them across the slit
during the integration to create a symmetric net slit function for the resulting spectra.
By comparison of multiple spectra
obtained of several Carina giants as well as by comparison of results from 
cross-correlation of different echelle orders, we find the random errors in the derived 
RVs to be better than 1.0 km s$^{-1}$ for the January run and 2.5 km s$^{-1}$ for the
December run.  The degradation in the second run was due to significantly
worse overall observing conditions
that resulted in poorer $S/N$ spectra on average. 

Table 1 gives for the stars observed with MIKE 
the J2000.0 positions, date of spectroscopic observation,
photometric data, RVs in both the heliocentric and Galactic standard of rest 
($v_{GSR}$) conventions, 
as well as 
a parameter that characterizes the quality of the RV:
an overall quality index, $Q$, which ranges from 1 (lowest quality) to 7 (highest quality).
The precise meaning of the various $Q$ grades is explained in 
\citet{kunkel1997a} and \citet{Majewski2004a}.

As an additional check on the RVs, 
we independently derived RVs for all MIKE spectra using the {\tt fxcor} package from 
IRAF\footnote{IRAF is
distributed by the National Optical Astronomy Observatories, which are
operated by the Association of Universities for Research in Astronomy,
Inc., under cooperative agreement with the National Science Foundation.} following
the method described in \citet{Frinchaboy2006}.  
The mean RV difference between both methods is
$0.1\pm0.4$ km s$^{-1}$ with a dispersion of $2.8\pm0.3$ km s$^{-1}$ showing a 
very close correspondence between the methods. 
However, for some spectra with very low $S/N$ our standard methodology failed to yield
an acceptable (i.e. $Q \ge 4$) cross-correlation, 
whereas {\tt fxcor} yielded a cross-correlation with higher apparent reliability.  
In these cases we have adopted the fxcor RV in Table 1 and given the derived velocity
error in place of the $Q$ value.

In the end, all 65 Carina-giant candidates observed with MIKE have a
reliable velocity and these form the basis of most of the outer Carina RV analyses below.
Among the 12 stars with RVs
selected from the expanded giant selection criterion in Figure
3b, none have been found to have a Carina-like velocity;
however, two of these stars have  
velocites near $v_{hel} \sim 332$ km s$^{-1}$ and constitute members of the
newly discovered halo substructure discussed in \S6.  Thus we include these two stars in
our discussions relevant to this halo substructure.

\subsection{The GIRAFFE Spectra}

Because  our  MIKE observing  focused  primarily  on  the most  widely
separated  Carina   giant  candidates,  our   resulting  spectroscopic
coverage  leaves a  significant statistical  gap from  the  only other
previously published  echelle resolution Carina RVs, which  are in the
Carina  core  (\citealt{Mateo1993}).    Fortunately,  there  exists  a
substantial  collection  of archived  VLT/FLAMES  observations of  the
Carina system  that bridges the  gap.\footnote{The archived FLAMES/VLT
data set used in this paper is part of the ESO large program 171.B-0520
``Towards  the Temperature  of Cold  Dark Matter:  Quantitative Stellar
Kinematics  in  dSph Galaxies",  PI.  G.  Gilmore.}   These data  were
retrieved and reduced to RVs by S.Z. and D.C..

FLAMES is installed at the Nasmyth A focus of the VLT Kueyen telescope
and is composed of a fiber positioner, OzPoz, that feeds the dedicated
medium-high resolution GIRAFFE  (resolving power $R\simeq 6000-30000$)
and UVES ($R\sim 40000$) spectrographs  with 132 and 8 science fibers,
respectively,  over a  large  field of  view ( $\simeq28$~arcmin  in
diameter) in the ``MEDUSA" mode.
The VLT Carina data set used in this
paper was collected over a 9 night run at the end of 2003 (22-31
December) and consists of 
16 different pointings, each observed four times. The exposure time
was $4\times3300$ seconds per pointing.
The four exposures for each pointing were taken in sequence and with the same MEDUSA plate
configuration.  All observations were done using GIRAFFE in the low
resolution, LR08 set-up having $R\simeq  6500$, and centered on the Calcium
infrared triplet to cover the region from 8206 to 9400 \AA\ .   At the
end of each observing night,  during daylight, a sample of calibration
frames were taken by the VLT staff within the nominal VLT calibration plan.

Spectroscopic calibration and extraction have been performed with the  
GIRAFFE  BLDRS\footnote{The GIRAFFE  BLDRS,  Base Line  Data
Reduction  Software, is  a  set of  python  scripts, modules  and a  C
library to reduce GIRAFFE  spectra. The software and documentation can
be found at http://girbldrs.sourceforge.net/.} data reduction pipeline 
(version 1.12)
and  the  GIRCALIB calibration reference file
database (version 2.3).  
The GIRCALIB image database contains
generic reference solutions for the calibration frames
(bias and dark frames, flat-fields, fiber slit geometry and fiber
response correction frames as well as wavelength calibration for all the different
FLAMES observing modes)
that are used as initial guesses for specific night-to-night
solutions for all the different calibration steps. 
For each night, the calibration frames (bias, flats, wavelength calibrations) are 
grouped together and reduced with the appropriate recipe, starting with the reference
solution in the GIRCALIB database and then iterating corrections to it.
Once all of the solutions are found they are applied with a single command
to the science images.

No particular problems were encountered in the reduction of the
calibration frames, 
but
occasionally the wavelength calibration gave unstable and distorted
solutions due to the uneven spacing and scarce number of ThAr
calibration lines in the available spectral range. To overcome and check this
problem, for each wavelength calibration frame used (one per night
taken during the daytime) the emission line detection threshold and
fitted polynomial order were readjusted  until a satisfactory
solution was obtained. These solutions were then verified directly by
the ThAr-calibrated science spectra (before night sky subtraction), which 
were cross-correlated  with a
separate,  emission line night sky spectrum calibrated
externally with the detailed night sky line lists of
\citet{Osterbrock1996,Osterbrock1997}.  It was found that the average
RMS velocity scatter from fiber to fiber based on the sky-lines was an
acceptable  $0.87$~km s$^{-1}$.  We adopt this value as our wavelength zero
point error for the GIRAFFE spectra.  The same test revealed that the offset 
from  plate to plate was less than $0.2$~km/s; nevertheless, we corrected all plates to
the same radial velocity zero point system based on the
night sky lines.

The archived GIRAFFE images contain spectra from all of the fibers for
a given MEDUSA plate.  Between 109 and 112 MEDUSA fibers were placed
on target stars depending on the pointing, with  the remaining fibers
positioned on empty sky positions.  The identification of the target
objects associated with each spectra is possible using associated
archived tables containing the observers' input values of target
positions and magnitudes as well as details of the positioning of the
fiber on the sky.

Radial velocity derivations were  performed using an implementation of
the  Tonry and  Davis  (1979)  method in  the  MIDAS environment.   We
extracted  radial velocities  both for  each single  exposure  of each
medusa plate and then for the sum of the four exposures per plate.   For
each  exposure we  first  extracted the  sky  fibers to  create a  sky
spectrum for that exposure.  This sky spectrum was subtracted from each
target fiber spectrum  and the result was continuum-normalized and
finally cross-correlated  with a synthetic  spectrum\footnote{We built the
template spectrum  using the Kurucz models properly  simulated for the
 GIRAFFE  spectrum resolution  and set-up used.   We  tested  several
templates   and finally  adopted  a  spectrum  for a  star  with
$T_{eff}=4500$, log$g=2.5$ and  [Fe/H]$=-1.5$.} of a low metallicity
giant star  to obtain the radial velocity.  In the second reduction
method  we summed the four extracted  and sky-subtracted spectra for
each star and cross-correlated  {\it that} with the template spectrum.
The comparison  of the single spectra and the summed spectra RVs for
each  object  revealed that the RVs from the former were very poor,
especially for the faintest stars: several times we failed completely
to measure a reliable RV. In the cases where we were able to get
four independent RVs we compared their average with the RV of the summed spectrum
and found that the RMS was much larger than the  measurement error in
80\% of  the sample.  Thus, we decided to use only the RVs derived from the summed
spectra.

A total of 1771 independent radial velocity measurements were obtained
across the sixteen 
medusa  pointings.  After removing 66 stars for which 
we could not get an RV 
and accounting for repeated targeting of some stars, RVs
were obtained for 994 distinct
stars.  In the final definition
of the  RVs we found that among
objects having more than three measurements
($\simeq130$ stars) the scatter was always compatible  with the
measurements errors except for the very faintest objects where we found
a larger scatter.  The RV errors take into account this larger scatter.
Of the 994 individual GIRAFFE target 
stars, 975 were found  in our Washington+$DDO51$  photometric catalog.
We only consider those 975 in our analysis because the remaining
19 stars not present in our photometric catalogues  cannot be checked
for their giant status in the 2CD.\footnote{The 19 stars missing from
our catalogue are primarily due to the loss of  stars in  the gaps
between CCD chips in our Mosaic images as well as to small gaps in the
placement of our Mosaic  pointings within $r_{lim}$, visible in Figure
1.  We note  that only 10 of  these 19 stars have  RVs consistent with
the  Carina dSph.}  Table 2  presents  the RV  information for  these
stars.

To  the  MIKE  and GIRAFFE  data  we  also  add Blanco+Hydra  RVs  for
photometrically selected giant  stars from \citeauthor*{Paper VI} that
were observed  at $R = 7600$  resolution in October  2001.  We include
these  Hydra  RVs only  for  those  stars  not already  having  higher
resolution echelle  observations.  In the  end, our sample  includes a
total of 1123 RVs  from Table 1, Table 2, \citealt{Mateo1993} and the
\citeauthor*{Paper VI} contribution.

\section{Spatial and Radial Velocity Distributions of Carina dSph Members } 

\subsection{Definition of Carina dSph Members}

\subsubsection{The Full Sample}

Figure 5a shows the distribution of all derived RVs for stars in the Carina field as 
a function of elliptical distance from the center, including stars having RVs 
from \citet{Mateo1993} (green points), 
Hydra observed stars from \citeauthor*{Paper VI} (cyan points), stars with 
MIKE RVs (red points), and GIRAFFE data (blue points). 
The elliptical radius of a star is defined to be the semi-major axis radius of the ellipse
centered on Carina (with the ellipticity, center and position angle for the dSph as found by
\citeauthor*{IH95}) that passes through the star.  Figure 6a shows the integral of 
the RV distribution over all radius.  A most obvious characteristic of these ``full sample" 
RV distributions is the presence of the
prominent RV peak associated with the Carina core near $v_{hel} \sim 220$ km s$^{-1}$.  However,
a significant contribution of stars at other RVs may be seen, particularly from stars with
$v_{hel} \lesssim 150$ km s$^{-1}$ from the Milky Way.  These contaminants 
come predominantly from the GIRAFFE sample, which was apparently primarily selected on the basis
of positions of stars in the CMD.  While the Carina RV peak still stands out, the substantial 
background of non-Carina stars makes it difficult to define an accurate 
RV criterion for cleanly isolating Carina members.

\subsubsection{The Conservative Sample}

Figures 5b and 6b show the same RV distribution, but only for stars satisfying the 
conservative Figure 3 
criteria for identifying Carina giant candidates by their Washington $M,T_2+DDO51$ photometry.
This distribution of the ``best" photometric candidates makes it easier to define an appropriate
additional criterion, based on RVs, for identifying Carina members.
Anticipating that the velocity dispersion of Carina members actually rises slightly 
outside the Carina core, we define as an RV membership criterion the 3$\sigma$ range 
defined by RVs for Carina
stars beyond $r_e>0.6 r_{lim}$ (twice the core radius), 
where we find a mean $v_{hel} = 220.8\pm1.3$ and 
a $\sigma=10.2$ km s$^{-1}$.\footnote{The velocity dispersions shown later in Fig. 11 
are at lower values than the
observed spreads in Fig. 5 because the former have been corrected for measurement errors.}
This range is indicated by the dashed lines in Figure 5b and the shaded region in Figure 6b.
This new RV selection criterion for MIKE and GIRAFFE stars is narrower than that applied in 
\citeauthor*{Paper VI}, but this is because the RVs in the present sample have 
smaller random errors. 
This final set of stars, selected by our conservative CMD and 2CD criteria (Fig. 3) 
is shown in Fig. 5b.

An additional feature apparent within the RV distribution of the ``best photometric sample" in 
Figure 5b and 6b is the distinct group of stars with a clumped RV at an even {\it more} extreme
velocity than the Carina dSph.  This feature is even more clear in Figure 6d, where we show a 
histogram for a subsample of stars from Figure 6b, in particular, stars 
with $r_e>1.5r_{lim}$. Their
$v_{hel} \sim 332$ km s$^{-1}$ implies a significant retrograde motion for stars
in this direction of the sky.  
The magnitudes and colors of these stars (Table 1) are also rather clumped, indicating
similar spectral characteristics and an apparently similar (and substantial)
distance. 
In \S6 we explore further this moving group of giant stars from what appears to be
a newly found halo substructure.

\subsubsection{Expanding the Conservative Sample}

A comparison of Figures 5a and 6a with Figures 5b and 6b suggests that the restrictiveness
of our ``conservative" photometric selection of Carina giants, while providing extremely
pure samples of Carina stars, also leads to a non-negligible
level of incompleteness (a well-known issue we have addressed before in \S3.2 of Paper VI).
Given that we now have the advantage of three criteria for discriminating Carina giants 
and a large number of RVs from GIRAFFE in the Carina main body, it
is worth reinvestigating the tradeoffs between sample size/completeness and 
sample purity.  More specifically, can we expand any of the selection limits
to admit substantially more Carina stars from the GIRAFFE sample 
without sacrificing the reliability of the membership census.  

{\bf 2CD outliers:}
Figure 7 demonstrates some possibilities for expanding our membership acceptance criteria
by showing the 2CD and CMD of stars satisfying our
newly established Carina RV-membership criterion, but falling outside one or the other (or both) our 
conservative photometric criteria (plotted as the solid lines in Figs. 3 and 7).  
As may be seen in Figure 7b, 
a large fraction of these stars lie {\it just below} our Figure 3b giant selection in the 
2CD.  However, inspection of the distribution of stars in Figure 3b clearly shows a strong, 
almost vertical giant star 2CD concentration at $(M-T_2)_0 \sim 1.2$ that extends below the adopted 
diagonal limit there.  Moreover, the 2CD analysis of giant and dwarf stars presented in \citeauthor*{Paper I}
makes clear that giant stars are commonly found at these positions of the CMD --- a point 
demonstrated by the superposition of the \citeauthor*{Paper I} ``giant star boundary" in Figure 7b
(dotted lines).  Stars in Figure 7 lying within the \citeauthor*{Paper I} 2CD boundary 
but within the Figure 3a CMD boundary are marked with red open triangles in Figure 7.  
Given that these stars satisfy the RV, CMD and the \citeauthor*{Paper I} 2CD 
criteria, we regard these stars as Carina giants from here on.

{\bf CMD outliers:}
We may also investigate those stars that satisfy the RV and 2CD criteria but not our initial
CMD boundary.  In Figure 7 these stars are marked with open blue circles for the stars
that satisfy the stricter of the 2CD boundaries and green open circles for the stars
satisfying the \citeauthor*{Paper I} 2CD limit.  Almost all of these lie very close to the 
RGB limit. A number of them lie 
at a CMD position just above the strong red clump.  Given that Carina
has stellar populations as young as 0.6-1.0 Gyr (\citealt{monelli2003}), it might not be too surprising
to find some core He-burning stars lying above the canonical red clump from the dominant, older,
more metal-poor Carina population (e.g. \citealt{salaris2002}).  Other modest CMD outliers 
are in CMD positions consistent with those expected for asymptotic giant branch stars.
A similar outlier trend was found in Paper VI, where it was noted that slightly expanding the magnitude
width of the CMD selection criterion by a few tenths of a magnitude would increase completeness with
virtually no decrease in reliability.  Given that previous conclusion, and that these stars satisfy
the 2CD and RV criteria, we consider all of these outliers as Carina members.

{\bf RV outliers:}
Finally, what about stars that fall within the 2CD and CMD criteria but just outside the RV criterion?
Several of these stars are conspicuous in Figure 5b.  First, we note 
that there are $\sim300$ stars satisfying
our 3$\sigma$ RV criterion in Figure 5b.  For a sample of this size and with a Gaussian distribution,
one expects $\sim 0.3$\%, or $\sim1$ outlier.  
As may be seen in Figure 5b, two stars with $r_e/r_{lim} < 0.6$
lie just below the RV cutoff and are probably very likely this kind of Gaussian-wing outlier member.
These stars ( C2661 and C161179)
are indicated by the blue solid squares symbols in Figures 5 and 
7, where they can be seen to be very solidly photometric members. 
Nevertheless, because they are in the well-populated central part of Carina,
whether or not they are included in our analyses
has very little effect. 

On the other hand, as we discuss in \S4.5, the velocity dispersion of 
Carina appears to grow beyond the King limiting radius, and, even though our RV selection 
criterion was derived from stars with $r_e/r_{lim} > 0.6$ specifically for this reason, 
the RV dispersion that sets the selection criterion 
is dominated by stars with $0.6 < r_e/r_{lim} < 1.5$.  Beyond this
range, the dispersion not only grows, but, as we show in \S5, the RV distribution becomes
flatter than Gaussian.  Both larger velocity dispersions as well as more platykurtic 
velocity distributions are fully consistent with models of disrupting dSphs 
systems (\citealt{Read2005a}; M06). 
Thus, even wider separated RV-outliers are not only conceivable at large 
radii, they are expected.  We mark three of these from our MIKE sample
--- C1960448, C2450090 and C2050415 ---
with red square symbols in Figures 5b and 7.  These stars, which lie within 
$\sim 28$ km s$^{-1}$ (3$\sigma$), $\sim20$ km s$^{-1}$ (2$\sigma$) 
and $\sim10$ km s$^{-1}$ (1$\sigma$), respectively, of our Carina RV membership limit,
are again solidly within the photometric
Carina giant candidate selection criteria (Figure 7).  They are particularly interesting potential
members, since all three lie approximately along the Carina major axis, and at large radii ---
$\sim 2.0 \deg$ to the east, $\sim1.6\deg$ southwest and $\sim2.0\deg$ northeast of Carina center, 
respectively (see Fig. 8a).  Indeed, the latter star is potentially the most widely separated Carina giant
in our sample, at $r_e = 4.9 r_{lim}$.  

Nevertheless, unlike in the cases of the sample-admitted 2CD and CMD outliers above,
even though we can make a compelling case for the membership of all five of
these RV outliers, we {\it exclude them} from our dynamical analyses to follow, so that we
do not unduly bias our velocity results.
Figures 5c and 6c summarize the RV distributions 
of our final, expanded Carina-member sample based on our two (slightly widened) photometric
criteria and one velocity criterion.  In Tables 1 and 2 we designate by the column ``Member" 
those 260 
stars considered to be members by the most conservative criteria and those 
additional 116 stars
that have been admitted as Carina members by the exceptions described in this subsection.  
We stress that (1) all 116 of these stars are from the GIRAFFE sample, 
(2) all but 2 are within $r_e < 0.9 r_{lim}$ and so have no 
impact on the dynamical results at larger radii, and (3) the inclusion
or exclusion of these 116 stars in our analysis has little effect on 
the general dispersion trends described later (Fig. 11).  Thus we have opted to 
include these 116 stars to improve our sampling and statistical 
uncertainties.
The five RV outliers discussed above but not included in our
analyses are highlighted in this column by ``RV?".

\subsection{Sky Distribution of Carina dSph Members}

The azimuthal distribution of the Carina RV-members on the sky  
(Fig. 8a) shows them to 
lie predominantly along the Carina major axis, even though, as
shown in Figure 1,
the azimuthal coverage of our photometric and spectroscopic efforts actually favors the
{\it minor} axes (see, e.g., the distribution of Carina giant candidates {\it not}
found to be RV members in Fig. 8b).
Figure 9, which shows the ratio of the circular to elliptical radius ($r_c/r_e$)
for each star in the survey versus its circular radius, demonstrates the tendency for Carina RV 
members outside the King limiting radius to lie along an extension
of the position angle of Carina's ellipticity and, indeed, to have an apparently 
even more elliptical distribution in this direction at larger radii.
Stars on the major axis
will have $r_c/r_e = 1$ and stars on the minor axis will have $r_c/r_e = 0.67$, according to 
the ellipticity of Carina (\citeauthor*{IH95}).
That the mean $r_c/r_e$ increases at larger $r_c$ shows the tendency for the extended population
to become even more stretched along the major axis,
evokes the character expected of tidal tails, and is a key characteristic of dSph
tidal disruption models
(\citealt{Oh95}; \citealt{PP95}; \citealt{JCG02}; \citealt{Choi2002}; M06).  
Further surveying 
for Carina members over larger radii to see 
whether and how this trend may continue would provide a valuable check and important 
constraint on the nature of any tidal disruption.

\subsection{Photometric Contamination Levels Revisited}

\citeauthor*{Paper VI} has already focused on 
the reliability of our methodology to 
assess dSph structure into extremely low surface brightness regimes, with specific focus on Carina.
However, with the now much better spectroscopic coverage 
as well as better photometry of the Carina field
we may reassess the effectiveness of our Washington $M,T_2,DDO51$ survey strategy. 
In addition, the MIKE spectroscopic sample, which was pre-selected based on the
Washington+DDO51 photometry, provides an interesting contrast with the GIRAFFE sample, which
was not.

A straight calculation of our success rate from the 48 Carina RV-members
among all 65 Carina giant candidates with MIKE spectroscopy yields a success rate
of 74\% in identifying true dSph members.  Restricting the analysis to only
stars outside the nominal (IH95) King limiting radius yields a success rate of 55\% (22 dSph
members among 40 $r_e>r_{lim}$ Carina giant candidates with RVs),  and this includes
candidates at extremely low densities (0.058\% the density of the Carina core).  
However, 13 of the 40 Carina giant candidates outside $r_{lim}$
with determined RVs appear to be giant stars
from {\it another} tidal stream with rather similar CMD characteristics as Carina (\S6).
Though these stars are not attributable to Carina, this newly discovered Milky Way 
feature might be argued as a success of the overall methodology we have been using 
in this series of papers to identify just this kind of halo substructure.  
Were we to combine these stars with the true Carina dSph members, our 
success rate in identifying ``halo substructure" stars rises to 94\%.

In contrast, the original GIRAFFE sample was apparently selected only on the basis of the position of 
these stars in the CMD (though not our CMD).  Among the 975 stars in the GIRAFFE sample
also in our catalogue, 
390, or 40.0\%, are found to have Carina RVs --- and this is for a sample highly concentrated to 
the main body of Carina, with most stars having $r_e < 1.0 r_{lim}$.
However, had we applied our photometric selection criteria to the GIRAFFE catalog
97.3\% of the stars identified as Carina giant candidates would have been found to be RV-members
(almost tripling the telescope efficiency).
Combining all available RV data at all radii, the Washington+DDO51 pre-selection 
results in a 90.5\% RV-member efficiency.  
Thus, the combination of Washington$+DDO51$ photometry with quality spectroscopy  
is found once again (see Palma et al. 2003, Westfall et al. 2006, Sohn et al. 2006)
to be a very effective observational strategy for identifying very diffuse halo substructures.

The point is relevant to potential further work on the extended structure of the Carina system.  Continued
searches for Carina giants at large separations from the dSph center
will require an efficient means to identify the best candidates to
optimally take advantage of spectroscopic time on the
largest telescopes.  We note that using {\it only} a selection for Carina
stars by their position along the Carina RGB in the CMD becomes a very 
inefficient way to find Carina giants at 3$r_{lim}$:
At these radii, only one in 85 stars in the RGB selection region in the CMD
we have used (Fig.\ 3b) turns out to be an actual Carina giant, and to $M = 20.8$, the
density of such stars is only 7.4 deg$^{-2}$, making even multifiber spectroscopic
searches for members within a CMD-only target list a rather inefficient enterprise.

\subsection{Standard Mass-to-Light Determination Revisited}

Estimates for the central and global Carina $M/L$ determined using standard prescriptions
(e.g., core-fitting combined with the central velocity dispersion) are given by Mateo et al. (1993) as
$(M/L)_{\rm o} = 40\pm23$ and $(M/L)_{tot} = 37\pm20$ (all $M/L$ values in solar units), 
respectively, when isotropic,
single-component \citep{King1966} models are adopted; anisotropic models were argued to 
give similar global $M/L$ for the lowest possible central mass density.  
These values were based on
an observed central velocity dispersion of $6.8\pm1.6$ km s$^{-1}$.  Monte Carlo analyses
conducted \citet{Mateo1993} show that it is unlikely that this dispersion has been inflated by
either atmospheric jitter in the target K giants or the influence of binaries.

However, there seems to be no real consensus on derived $M/L$'s for Carina.
For example, Mateo (1998) quotes the Carina $(M/L)_{tot}$ as 31, 
whereas \citeauthor*{IH95}, adopting
the original Mateo et al. (1993) central velocity dispersion, 
derive $(M/L)_{tot} = 59\pm47$ and $(M/L)_{\rm o} = 70\pm50$ (where the large
error bars reflect uncertainties in the velocity dispersion, core radius and 
at least a factor of two uncertainty for the central surface brightness).  
\citet{Walcher2003} estimate the Carina mass and $M/L$
by assuming that its periGalactic tidal radius can be 
approximated by $r_{lim}$ (obtained from their photometric survey of the dSph)
and using the \citet{Oh1992} relationship between 
the tidal radius of a satellite and its mass and orbit. 
Circular orbits yield
$M/L$ as low as 0.6 while more eccentric orbits can easily accommodate values as high as the ones
derived by \citet{Mateo1993}, but \citet{Walcher2003} derive a Carina $(M/L)_{best} = 17$ based on an orbit with 
eccentricity 0.6 and apoGalacticon twice that of Carina's current distance.

The new RV dataset presented here invites yet another $M/L$ evaluation.  
Unlike previous determinations making use of a ``central" velocity dispersion from a relatively
small number of stars in the very core of the dSph, our extensive and 
radially continuous velocity coverage means that the definition of "central" is not pre-defined
by our available sample.
If we assume that at least the inner parts of the dSph are well represented by a
King profile, Figure 4.11 from \citet{BT1987} shows that the velocity dispersion of
stars begins to deviate from its central value at about half the core radius. 
Figure 10 shows the central velocity dispersion of Carina as we grow the radius (shown in units
of core radius as measured by \citeauthor*{IH95}) within which we 
include RVs in the dispersion computation.
As we add successive stars out from the Carina center
the derived ``central" velocity dispersion (calculated using the maximum likelihood
method, \citealt{Pryor1993}; \citealt{Hargreaves1994}; \citealt{Kleyna2002})   
reaches a value of $6.97\pm0.65$ km s$^{-1}$ at half
the core radius (computed from 87 total Carina stars).  
This value, which is slightly larger than (but consistent with) the 6.8 km s$^{-1}$ value 
used by \citet{Mateo1993}, is adopted to rederive the Carina $M/L$'s. 

The central mass-to-light ratio can be determined as (\citealt{Richstone1986}):
\begin{eqnarray}
(M/L)_{\rm o} = {\rho_{\rm o} \over {I_{\rm o}}} = \eta {333 \sigma^{2}_{\rm o} \over {{r_{1/2} S_{\rm o}}}}
\end{eqnarray} 
where $\eta$ is a correction parameter dependent on the concentration value (0.955 for
Carina), $r_{1/2}$ is the geometrical mean of the half-light 
radii measured along the major and minor axis
($163\pm26$ pc)
and $S_{\rm o}$ is the central surface brightness ($2.2\pm1.0$ L$_{\sun}$/pc$^{2}$). 
We adopt all these structural values from \citeauthor*{IH95}\footnote{Aside from fitting the
presently derived Carina density distribution, these parameters also fit well the
Carina distributions in \citet{Walcher2003} and Paper II. 
Moreover, they fit our data better than the
parameters derived by \citet{Walcher2003} from the theoretical King 
model (\citealt{King1966}).}  
and obtain $(M/L)_{\rm o}=43^{+53}_{-19}$ for Carina where the main source of uncertainty
comes from the uncertainty in the central surface brightness. To illustrate this, we
calculate the error in the $(M/L)_{\rm o}$ not considering
the uncertainty in the central surface brightness, and
obtain $(M/L)_{\rm o}=43^{+8}_{-7}$.

From \citet{Illingworth76}:
\begin{eqnarray}
(M/L)_{\rm tot} = {{166.5 R_{c,g} \mu} \over {\beta L_{\rm tot,V}}}
\end{eqnarray}
where $R_{c,g}$ is now the geometric-mean King core radius in pc ($210\pm30$), $\mu$ is the
\citet{King1966} dimensionless mass parameter, and $\beta$ is a model-dependent 
velocity parameter related to the observed velocity dispersion.  Table 
10 in \citeauthor*{IH95} gives values for both $\mu$ and $\sqrt{\beta \sigma^{2}_{\rm o}}$ of
$2.8\pm1.3$ and 0.52, respectively, 
for a Carina concentration of $\log(r_{t}/r_{c}) = 0.52$.  This yields 
$(M/L)_{\rm tot,\rm V} = 41^{+40}_{-25}$ 
for a $L_{\rm tot,\rm V}=0.43\times10^{6}$ (\citealt{Mateo1998}).
This translates into a total mass of $M_{\rm tot}=1.76^{+1.75}_{-1.10}\times10^{7}$ M$_{\sun}$.
These results are in very good agreement with the ones found 
by \citet{Mateo1993} despite the fact that
the structural parameters they use are different from the IH95 ones adopted here. 
Here we adopt the updated distance of Carina from Mateo (1998), which is larger than
the value used by \citet{Mateo1993}, 
and this results in a larger half-light radius that  
compensates for the slightly larger luminosity adopted here.

\subsection{Velocity Dispersion Trend of Carina Stars}

With this large RV dataset in hand we can now assess the velocity dispersion
behavior for Carina to well past $r_{lim}$.
To ascertain this trend, we have studied the velocity dispersion as a
function of both elliptical and circular angular distance from the Carina center.  
Because the true shape of the gravitational potential and tidal 
boundary of a dSph are likely to be somewhere in between 
these limiting shapes, it is 
helpful to explore these two limiting cases.  In each calculation of an RV
dispersion 3-$\sigma$ outliers have been
removed iteratively, with the mean velocity for each bin reevaluated at each iteration 
and the dispersions estimated using the maximum
likelihood method. We note that this method assumes that the
velocity distribution follows a Gaussian distribution everywhere
which is not strictly true for Carina.
However, such non-Gaussian behavior is apparent only in the outskirts of 
Carina ($r \gtrsim r_{lim}$; \S5), and the effect of the non-Gaussian character 
found there is that the dispersion will tend to be slightly underestimated by 
the maximum likelihood method.

Figure 11 shows the derived Carina velocity dispersion profiles for both choices of angular 
separation:
the left panels show profiles plotted against elliptical radius, the
right shows the same for circular radius.  To test binning effects, we have used both 23 
and 46 stars per bin (lower and upper panels respectively) for stars 
inside $r_{lim}$, but because the number of stars
with measured RV beyond this point is sparse, the last four dispersion points in each plot 
are binned at 10 stars each.

The Figure 11 Carina profiles remain fairly flat throughout the radial extent of the 
main body of the dSph, to $\sim1.1r_{lim}$. 
Such flat profiles over a comparable structural
radial range have now been reported 
(although not to the radial extent of this study) for several 
dSphs: Sculptor (\citealt{Tolstoy2004}; \citealt{Westfall2006}), 
Draco (\citealt{Mu05}), Ursa Minor (\citealt{Mu05}),
Fornax (\citealt{Walker2005}), Leo I (\citealt{Sohn2006}) and 
Sagittarius (Majewski et al., in preparation). 
Note that while \citet{W04} found a sudden drop in velocity dispersion
at about $r_{lim}$ for both Ursa Minor and Draco, this feature could not be
reproduced by \citet{Mu05} when reanalysing these profiles when Washington+$DDO51$ 
photometric and additional spectroscopic data were used to check them.
Kleyna et al. (2004) have also found Sextans to have a predominantly flat profile but with a 
cold velocity dispersion at about $r_{lim}$ (and a kinematically cold center as well); given 
that similar claims for cold points near $r_{lim}$ in the Ursa Minor and Draco dSphs
have not held up under further scrutiny, the Sextans result warrants further investigation.

Flat velocity dispersion profiles are incompatabile with mass-follows-light dSph models (with
or without dark matter) in complete dynamical equilibrium, where decreasing dispersions are
expected at large radius, approaching zero as the cutoff radius of the distribution is approached.
To explain the observed velocity behavior,
\citet{Walker2005} suggest that the {\it easiest} assumption to discard is that mass follows light;
following this line of reasoning, a number of 
groups  (e.g., \citealt{Lokas2005}; \citealt{xiao2005}; \citealt{Read2005b}; 
\citealt{Mashchenko2005b}; \citealt{Walker2005})
have invoked ``two-component dSph models",
where the dark mass extends far beyond its luminous counterpart
and is responsible for the flat dispersion profile at large radius.
Yet, our MIKE observations of Carina have now yielded the most extensive 
coverage of velocities in any dSph,
including, for the first time, the measurement of the
velocity dispersion of a dSph (apart from Sgr) with a reasonable sample of stars beyond $2 r_{lim}$.  
As may be seen in Figure 11, the velocity dispersion for Carina approximately doubles at these
large separations --- a result that is {\it not} explained with previous
two-component models.

Is abandoning mass-follows-light really the ``easiest" 
assumption to discard in the dSph models?
Flat dispersion profiles arise {\it naturally} in tidal disruption models
(\citealt{kuhn1989}; \citealt{Kroupa1997}; \citealt{fleck2003})
{\it even if large amounts of dark matter are present} and the central parts of dSphs are
bound and in equilibrium (\citealt{Mayer2002}; \citealt{Sohn2006}).
As we show in M06, a single-component, mass-follows-light,
tidally disrupting dSph model gives a good representation for both the density and velocity
dispersion profile for the Carina dSph we have derived here.

Further evidence for a disruption scenario is provided by the trend of
velocity across the satellite.
In Figure 12 we show the mean RV (in Galactic Standard of Rest)
as a function of $b$-distance from the center of Carina (approximately
the major axis of the satellite).
No significant RV trend in the central part of Carina
that resembles a rotation curve is observed.
However, beyond $r_{lim}$, a gentle velocity gradient 
is observed across the major axis of Carina to the extent of our observations.
Over $\sim1.2$ degree (2.1 kpc), a peak-to-peak difference of $\sim10$ km s$^{-1}$
is seen in this trend --- a difference
significantly larger than the error in the means for the binned points.
This velocity trend is interesting because 
it has been predicted as a hallmark of tidal disruption by several
studies (e.g., \citealt{PP95}; \citealt{JSH99}, \citealt{fleck2003}).  According
to \citet{Pryor96}, ``a velocity gradient across the galaxy that is larger than the velocity
dispersion is the clearest signature [of tidal destruction]".

\subsection{Implications of Widely Separated RV-Members}

Figures 5 and 8 show that we have found RV-verified Carina member stars to 4.5 $r_{lim}$.  This
limit may extend to 4.9 $r_{lim}$ if we adopt a 3$\sigma$ limit for RV-members specific to the
outermost bins in Figure 11, in which case star C2050415 (represented by the outermost square
in Figures 5b and 8) is the outermost detected Carina giant.
If the RV member at 4.5 $r_{lim}$ is bound to Carina, it sets a new lower limit for
the physical extent and tidal radius of the dSph at 96.5 arcmin, or 2.84 kpc for an assumed distance
of 101 kpc to Carina (\citealt{Mateo1998}).
Using this radius in the 
tidal limit equation (\citealt{Oh1992}): 

\begin{equation}
R_{tidal} = a \left( {M_{\rm dSph}\over {M_{\rm G}}} \right)^{1/3} \left  \{ {(1-e)^2 \over {[(1+e)^2/2e]{\rm ln}[(1+e)/(1-e)]+1}} \right \}^{1/3}
\end{equation}

\noindent where $a$ is the orbital semimajor axis, $M_{\rm dSph}$ and $M_{\rm G}$ are the mass of the
dSph and the MW inside $a$ respectively and $e$ is the orbital eccentricity (values for $a$ and
$e$ taken from \citealt{Piatek2003} to be 61 kpc and 0.67 respectively),
the lower limit to the Carina mass becomes 
$2.7\times10^9$ M$_{\sun}$ assuming a mass of the Milky Way interior to $a$ of
$M_{MW}=6.7\times10^{11}$ M$_{\sun}$
(\citealt{Burkert1997}).  This estimated mass limit is further underestimated 
because we are taking the {\it projected} radius of the star as the actual, three-dimensional
distance from the center.
Given the Carina luminosity $L=0.43\times10^6$ L$_{\sun}$ (\citealt{Mateo1998}), 
the above mass translates to a global mass-to-light of $M/L > 6,300$, which is more 
than 100 times higher 
than the central and total $M/L$ derived for Carina in \S4.4
\footnote{These estimations are robust to the uncertainties in the orbital parameters
derived by \citet{Piatek2003}. Their 95\% confidence range for $e$ is (0.26; 0.94)
which results in a $M/L$ range of (370; 470,000). Even a value for $e$ of 0.24 
corresponding to an orbit with peri:apoGalacticon of 63:102 kpc, (their 95\%
confidence bounds for these parameters) yields a
$M/L$ that is an order of magnitude higher than the central value.}.
On the other hand, if the star at 4.9 $r_{lim}$ is a Carina member and it is bound, it sets
the tidal radius at 133.7 arcmin, or 3.93 kpc, enclosing an astounding mass 
of $7.2\times10^9$ M$_{\sun}$, which yields $M/L > 16,000$.

While some stars on trapped orbits can be found well outside the true tidal radius 
up to 2$r_{lim}$ or even more 
(see, e.g., discussion in \S7.3 of \citealt{BT1987}), the number should 
be extremely rare beyond 4$r_{lim}$.  Also, were one to expect the $M/L$ of a galaxy
to grow with radius, the asymptotic values implied for Carina are unreasonable high 
even when compared to values for galaxy clusters: 200 - 300 (\citealt{Carlberg1997}), 
which are thought to be approaching fair samples of the universe.
From this line of reasoning, we must therefore conclude that either Carina has an 
enormous, extended dark matter halo to create a $M/L$ an order of magnitude higher
than the universe, or, more simply, that these widely separated Carina stars are 
simply not bound.

We (Mu\~noz et al. 2005) have used similar arguments in our discussion of the Ursa Minor dSph, where 
a global $M/L$ of 1,400 to 14,400 was implied by the widest separated RV member, 
depending on the use of circular or elliptical radii, respectively.
While the possibility that the widely separated Mu\~noz et al. Ursa Minor stars 
could be interlopers that just happen to have
the same RV and color-magnitude positions (i.e. approximate distances) as Ursa Minor
was explored and shown to be very unlikely, this miniscule possibility 
cannot presently be completely discounted.  However, the case for the widely 
separated Carina stars being interlopers is far 
more difficult to make because of the sheer number of them: 
six (possibly eight) farther than 2$r_{lim}$.  
Figure 13, which shows the global mass and $M/L$ implied for Carina as progressively more 
widely separated RV members are attributed as bound satellite members, demonstrates
that the implication of an enormous implied Carina $M/L$ is robust 
to the invalidation of any 
particular star, or even several, attributed as a sample interloper.
The $M/L$'s in Figure 13 are derived in two ways that make use of equation 3:
(1) The implied mass of Carina is found by assuming a spherical 
potential for the dSph and the star's linear
projected distance from the center of Carina used as $R_{tidal}$ ({\it open circles}; again, this is a 
conservative lower limit, because we are working with {\it projected} radii).  
(2)  Assuming that the distribution
of stars around Carina maintains a constant ellipticity with radius, we can assume 
there exists for every star not on the major axis a counterpart at the 
same {\it elliptical radius} on the major axis which is then used for $R_{tidal}$.  
This assumption raises the lower limits on the implied $M/L$'s
({\it solid circles}).  The two methods for deriving the minimum implied $M/L$ probably 
span the actual limits, since galaxy potentials tend to be rounder than their 
density profiles.  

Figure 13 demonstrates that all of the stars with $r_e$ or $R$ exceeding $0.8 r_{lim}$ 
would need to be discounted as Carina-associated to bring the global minimum $M/L$ to 
more standard values for the Carina dSph (such as the $M/L \sim 40$ found from 
core fitting with the central velocity dispersion in \S4.4). In other words, if one
assumes that the global $M/L$ of Carina is that obtained using the central velocity
dispersion, then the tidal boundary {\it coincides}
with the radius at which the break in the density distribution is indeed observed.

Figure 5b attests to the
relative purity of the Carina dSph giant candidate sample created by our dual photometric
selection criteria (Figs.\ 3a and 3b):  Very few RV outliers are found among our Carina giant candidates 
overall, and, in addition 
the small number of giant candidates we find that do {\it not} share the Carina dSph RV 
lie predominantly in the 332 km s$^{-1}$ group.
Furthermore, Figure 5b suggests that the outer halo 
is highly substructured (at least when traced by giant stars), 
a result that is also evident from Figure 2 in Mu\~noz et al. (2005). 
In such circumstances, to obtain substantial contamination in our survey
would require a considerably
unfortunate conspiracy of phenomena to produce a {\it second} halo substructure with
the same RV, approximate distance, and CMD distribution as Carina; we consider
this possibility as unlikely.

\section{The Case for Tidal Disruption of the Carina dSph}

Taken alone, Figure 13 can be argued as a validation of the notion that dSphs
like Carina are surrounded by large dark matter halos (\citealt{Stoehr2002}; \citealt{Hayashi2003}). 
According to \citet{Hayashi2003}, NFW-like halos that fit the Carina
central velocity dispersion (adopted as 6.8 km s$^{-1}$) and central luminous King
profile, even in the face of substantial tidal stripping of the dark halo, still maintain
halos with (1) maxima in their circular velocity profile exceeding 50 km s$^{-1}$ that peak
well outside $r_{lim}$, as well as (2) 
true tidal radii of 11 kpc or more.
Making similar arguments for all of the Milky Way 
satellites alleviates --- {\it at the high mass end} ---
the mismatch between the CDM-predicted subhalo mass function
and that presented by the Galactic satellite system (i.e., the ``missing satellites
problem"; \citealt{Kauffmann1993}; \citealt{Klypin1999};
\citealt{Moore1999}).  

Nevertheless, we believe that an alternative explanation of 
Figure 13 --- i.e. that Carina (and other dSphs)
are surrounded by populations of {\it unbound} stars released 
through tidal disruption --- is not only simpler but also provides a
better match to {\it all} of the available observations of Carina:

{\bf Density profile}: We have remeasured the Carina density profile with new data, 
and confirm
the existence of a two-component, ``King+power law break" shape suggested earlier
by the photometric studies of \citeauthor*{IH95}, \citet{kuhn96}, 
\citeauthor*{Paper II},
and \citet{monelli2003, monelli04}.  This photometric work is now solidly
backed by spectra of stars in the break population (see also \citeauthor*{Paper VI}), 
proving the existence of RV-members
in the extended power-law break population and leaving no doubt as to the reality 
of the feature
(cf. \citealt{mor01}; \citealt{Walcher2003}).
This density profile matches (1) the classic shape of a disrupting dSph galaxy, as seen by 
N-body simulations of disrupting satellites (e.g., \citealt{JSH99}, \citealt{Mayer2002}) 
as well as (2) profiles observed in archetype examples of tidal disriuption 
like the Sagittarius system (\citealt{MSWO}).
In contrast, no published dark halo models predict a dynamical structure
that would give rise to the observed {\it luminous}, two-component profile of Carina.
It is difficult to imagine how the required structural transition between two bound, pressure-supported
stellar populations\footnote{We find 
little evidence for rotation in either the King profile or power law components
of the structural profile of Carina within $r_{lim}$.}
could be produced so deeply
inside an extended dark matter halo, and, coincidentally, exhibit {\it no significant change} 
in the observed dynamics (velocity dispersion) at this point (see below).
Moreover, the position of the break in the profile precisely matches that 
expected for a Carina having a constant $M/L$ given by the core-fitting technique (\S4.4). 

{\bf Azimuthal configuration}: The distribution of stars found in the outer Carina structural component
shows a preference to lie along the major axis, and to have an even greater ellipticity
than the Carina core, just as would be expected for emerging tidal tails (e.g., 
\citealt{Oh95}; \citealt{PP95}; \citealt{JCG02}; \citealt{Choi2002}).  
In contrast, CDM halos tend to have rounder 
potentials (\citealt{Stoehr2002}; \citealt{Hayashi2003}; \citealt{bailyn2005})
so that either the Carina halo is very unusual, or an explanation is required for
why its embedded luminous component has a rather different spatial distribution than its 
dark halo.

{\bf Velocity shear}: As pointed out in \S4.5, the observed velocity trend observed in
the Carina system is that expected for tidally induced shear.
However, we regard this observed trend with caution appropiate to the still 
meager statistics for this measurement in the outermost parts of Carina.

{\bf Velocity dispersion profile}: We find a Carina velocity dispersion profile that is
flat and then rising well past the King limiting radius.  
A characteristic of bound
populations is that eventually the velocity dispersion of stars should decline with radius, 
eventually approaching 0 km s$^{-1}$ at radii where bound
stars reach the apocenters of their internal orbits.  
That a dynamical ``cold point" radius is
{\it not reached} even among our most widely separated RV-members suggests that, if
bound, these stars are not near the tops of their orbits, and that the tidal radius
of Carina must be beyond --- 
even {\it well beyond}, given the still large velocity dispersion at 
$\sim2.5$$r_{lim}$ --- the observed typical radius
of our RV-members.  Thus, to explain the observed velocity dispersion trend requires an extremely
extended dark halo of even larger dimensions and mass than implied by Figure 13.  

In contrast, flat (and rising) dispersion profiles are a natural product of
tidal disruption models (\citealt{Kroupa1997}, M06).

{\bf Flattening of the velocity distribution}: 
As shown in recent studies (\citealt{Mashchenko2005b};
\citealt{Walker2005}; M06) if the Milky Way tidal field strips stars from
dSphs (even if surrounded by a DM halo) the velocity distribution at
large radii deviates from a pure Gaussian, in general becoming more
platykurtic near and beyond $r_{lim}$. We have shown for the case of Ursa 
Minor, Draco (\citealt{Mu05}), Sculptor (\citealt{Westfall2006}) and 
Leo I (\citealt{Sohn2006}) that the velocity distribution evolves from 
Gaussian in the center to a flatter distribution with increasing radius.  
The same is observed in Carina, where the distribution seems to flatten out
at large radii, with a kurtosis excess of $\gamma_{2}=-0.9\pm0.6$ for stars 
beyond 0.8$r_{lim}$ contrasted with the near-Gaussian 
$\gamma_{2}=+0.2\pm0.2$ for stars inside 0.8$r_{lim}$.
However, we note that such flattened outer RV distributions could also be
observed in systems where the orbits are mostly 
circular (\citealt{Dejonghe1987}).

{\bf An emerging ``too many satellites problem"?}:  \S4.6 makes the case that to keep all of Carina
RV members bound requires a potential minimum mass for the dSph 
of $\sim1.0 \times 10^{9}$ M$_{\sun}$.
\citet{Mu05} have performed a similar analysis on the Ursa Minor dSph system and find 
that to keep it's most widely separated RV-member bound requires a minimum mass of
almost $10^9$ M$_{\sun}$, or $10^{10}$ M$_{\sun}$ for a counterpart of 
that star moved along its elliptical isopleth
to the major axis.  
\citet{Read2005b} argue that, in fact,  dSphs have masses of $10^9$ -- $10^{10}$
M$_{\sun}$, which would prevent them from undergoing tidal stripping, even
in very extreme, radial orbits.
Such $\sim$LMC-mass dark matter halos (DMH) are at the limits of 
the largest subhalo sizes predicted by $\Lambda$CDM (\citealt{Mashchenko2005a}); 
the existence of {\it several} $\sim LMC$-mass subhalos in a Milky Way-sized
system is not expected (see Figure 14 of \citealt{Hayashi2003}).
If more examples of subhalos much more massive than previously inferred are found --- e.g., if we
continue to extend the radius over which RV-members are identified in Carina and the other 
satellites of the Milky Way (see, e.g., \S6) and attribute these stars as bound to the dSph ---
a new problem for CDM will
emerge, namely an {\it excess} of inferred massive satellites about the Milky Way.  
While the situation is not yet extreme enough to rule out the extended dark halo hypothesis 
on this basis, nevertheless, it is worth pointing out again that tidal disruption is 
a simple way to put stars at any arbitrary angular separation from a dSph, should 
even more extreme outliers be found.  Moreover, as \citet{Read2005b}
point out, inferring the existence of these extremely extended halos and large masses for satellite galaxies brings an inconsistency with the actual measured central velocity
dispersions (which are lower than predicted), even if significant tidal 
stripping and shocking are considered. 

{\bf The Sagittarius paradigm}: All of the observed spatial and dynamical
features in Carina are also found in
the one undisputed case of dSph tidal disruption in the Milky Way --- the Sagittarius
dSph (see Sgr spatial and velocity properties given in \citealt{MSWO,Majewski2004a}).  
Moreover, we (M06) have 
explored N-body simulations of modest mass, one component dSph systems (originating
as Plummer models) orbiting for significant fractions of a Hubble time
and can reproduce the observed properties of Carina fairly well.  That {\it both} (1) 
an actual, uncontested,
{\it tidally disrupting} analogue of the Carina system, as well as (2) successful tidal
disruption models 
(with fewer unexplained details than alternative, extended dark matter halo models) exist makes 
it difficult to avoid the question: Is Carina simply another example of the established
Sgr paradigm?

{\bf Commonality of disruption}:
A number of discoveries of 
apparent halo moving groups or streams have recently been made (including the one presented here in the 
foreground of Carina, see \S6): 
the Monoceros/GASS stream (\citealt{Newberg2002}; \citealt{Ibata2003}; \citealt{RP2003};
\citealt{Crane2003}), the TriAnd structure (\citealt{RP2004}, Majewski et al. 2004), 
the M31 giant southern stream
(\citealt{Ibata2001}) and a recently discovered, second M31 halo substrucutre 
(\citealt{Kalirai2005}),
the identification of an outer Galactic halo stream using blue horizontal
branch stars by \citet{Clewley2005}, a potential system in Virgo (\citealt{Duffau2006});
and a new halo moving group found with M giant stars (Majewski et al., in preparation). 
This growing list of examples
provides increasingly solid evidence of a highly substructured Milky Way halo, and to 
the {\it commonality} of tidal disruption of stellar systems in the Milky Way halo 
(e.g., \citealt{Font2006}; \citealt{BJ2005}).  Such tidal streams must come from 
{\it somewhere} and dSph satellites are the most obvious available source.

\section{Discovery of a Dynamically Cold Moving Group in the Carina Foreground}

\subsection{Observed Properties of the 332 km s$^{-1}$ Group}

The new MIKE RVs have revealed an additional 
coherent RV peak in the field centered on the Carina dSph (Fig. 5)
at $v_{hel}=332.2\pm2.6$ km s$^{-1}$, represented by 15 stars with the
rather small velocity dispersion of $9.8\pm1.9$ km s$^{-1}$ (Figs.\ 5 and 6).  
The extreme RV of this system (+122 km s$^{-1}$ when converted to the 
Galactic Standard of Rest) implies a strong retrograde motion for these
stars if they are nominal Milky Way stars at this Galactic position ($[l,b]=[260,-22]^{\circ}$). 
The strong RV coherence of this group makes it even more unlikely
that it is from a dynamically hot, well-mixed, random Galactic halo population, but the dispersion
is, however, of order 
what one sees in dwarf satellite galaxies: 
For example, the dispersion is comparable to those measured in the extended 
parts of the Carina system (Fig.\ 11) --- which we have argued to be likely tidal
debris --- as well as those measured
all along the trailing tidal arm of the Sgr dwarf debris stream (\citealt{Majewski2004b}).
However, the lack of any spatial concentration of these stars across the relatively
large span of our survey fields (see
Fig.\ 8b) and their very low apparent density 
(a factor of $\sim2$ more diffuse than the mean $r > r_{lim}$
giant star density for Carina stars of the same apparent magnitude) suggest
that these stars represent either tidal debris from a satellite galaxy
or an extremely low density part of a very extended satellite.

Figure 14a shows the distribution of stars in this moving group within the 
CMD of all stars selected to be giants in our photometric survey  
(according to the tenets of Figure 3b), along with the ``Carina dSph RGB" boundary
we have used in Figure 3a.
The CMD positions of the fifteen 332 km s$^{-1}$ group stars is
both highly concentrated and
slightly brighter in mean RGB position than
the mean CMD locus of the Carina RGB.
A similar concentration is also seen for the  
moving group members in the 2CD (Fig.\ 14b)\footnote{Note that two of the fifteen 
moving group stars lie just outside our more conservative giant selection criteria, 
and were part of the experimental foray into this region with the MIKE sample
discussed in \S3.1.}; 
moreover, their relative position
in the 2CD compared to Carina stars suggests
that the 332 km s$^{-1}$ stars are more metal rich than the mean Carina star
(see \citeauthor*{Paper I}), assuming similar [Mg/Fe] ratios.

An independent test of the relative metallicities of these stars comes directly
from the spectra: Despite the relatively low $S/N$ of the spectra (which were taken
for RVs), in many cases the strong calcium infrared triplet lines are clear.  
When possible the equivalent width for each triplet line within each MIKE
spectrum was measured.  We found that for all three calcium lines the equivalent widths for the
332 km s$^{-1}$ group stars were about double those of Carina stars with a similar $(M-T_2)$
color.  

We also used a photometric bandpass method for measuring the calcium infrared
triplet line strengths because (1) it is perhaps more reliable for relatively low $S/N$ spectra, 
(2) it averages results over three lines, and (3) a formalism exists to convert these 
photometric line measures into 
a formal [Fe/H] value.  We limit this work to MIKE spectra with $S/N\ge7$ per pixel
and follow the bandpass definitions summarized in \citet{a88}. 
We point out that since our original survey was not intended to measure
metallicities, we did not observe an appropriate set of 
stellar calibrators of the metallicity scale.  
However, since a primary intention is to compare the relative metallicity
between the Carina and 332 km s$^{-1}$ group samples, precise calibration is not necessary.
Therefore, we followed the prescription outlined in \citet{cole04} for converting
calcium equivalent width and stellar gravity to [Fe/H], adopting the calibration for this procedure  
from \citet{Koch06}.
A. J. Cenarro graciously made available the code used to measure the
line strength indices (\citealt{cenarro1,cenarro2}).
For studies of resolved galaxies and star clusters an RGB star's CMD position relative to the
system horizontal branch, $V-V_{HB}$, is often used as a proxy for surface gravity. 
To adopt this method, transformation equations from Majewski et al. (2000a) are used to translate the
Washington photometry into Cousins $V$ and $I$ magnitudes.  We start by assuming
all stars are at the same distance as the Carina dSph and
adopt $V_{HB}$=20.8 as the mean magnitude of the Carina red horizontal branch.
\citet{Frinchaboy2005} use a similar technique to study open clusters with 
spectra having only slightly better $S/N$ and derive a mean metallicity error of 0.3 dex.
Therefore, we believe that 0.5 dex is a conservative estimate of our mean uncertainty, 
where the main contribution comes from uncertainties in the equivalent width measurements.

Figure 15 shows the [Fe/H] distribution derived for both Carina and 332 km s$^{-1}$ group stars 
under the assumption of a similar distance.
The mean [Fe/H] derived for
Carina stars is -1.86 with a dispersion of $\pm0.41$ --- in good agreement with other studies 
(\citealt{monelli2003}; \citealt{Koch06}) --- whereas
the mean [Fe/H] derived for the 332 km s$^{-1}$ group is -0.93 with a dispersion of 
$\pm0.62$. 
Barring possible variations in [Ca/Fe] between the two groups of stars, 
Figure 15 suggests that the metallicity of the 
moving group may be $\sim 0.9$ dex higher in [Fe/H] than the Carina dSph were this
group at the same distance.

\subsection{The Magellanic Cloud Connection}

On the other hand, if these moving group stars are more metal rich (as their calcium line
strengths suggest), they are also {\it intrinsically fainter} in the $V$ band, whereas 
they are also {\it brighter} in apparent magnitude
relative to Carina stars of the same color.  
All of this suggests that the moving group
must be {\it closer} than Carina, and by as much as 
a magnitude in distance modulus or more (see, e.g., Fig. 12a of \citeauthor*{Paper I}).  Interestingly,
this places the distance of these stars to be of order the distance of the 
Large Magellanic Cloud (LMC), {\it the center of which not only lies only $\sim20^{\circ}$ away 
from the center of our Carina field 
in the sky but has a similarly high systemic
heliocentric velocity (262 km s$^{-1}$; 
\citet[][hereafter {vdM02}]{vdM02}}.
Even more intriguing, in other $M,T_2, DDO51$ photometric survey
work in fields encircling the LMC 
we have found additional giant stars with LMC-like velocities
ranging from 4 to 18.5$^{\circ}$ away from the LMC center in the 
general region between the LMC and the Carina dSph.  
Preliminary results for this work have been shown in
e.g., Fig. 6 of Majewski (2004), and a more complete discussion
will be given elsewhere (Nidever et al., in preparation).  Here
we focus on the relative positions (Fig. 16) and velocities (Fig. 17) 
of our best-quality velocities for stars in fields that bridge the region between
the LMC core and our Carina survey field. 
Because the expanse of sky involved is
sufficiently large that there is significant variation in the reflex motion of the
Sun in the RV, Figure 17 shows velocities after conversion to the Galactic Standard of Rest
(GSR) frame.\footnote{Figure 17 shows {\it all} giant candidates in our survey regions with
measured RVs within the plotted $v_{GSR}$ range; groups of stars with clumped, 
negative (i.e. generally retrograde)
$v_{GSR}$ are also found (e.g., see \citealt{Majewski2004b}), but are not relevant to the 
present discussion.}
\footnote{The adopted motion of the Sun is (232,9,7) km s$^{-1}$ in the Galactic rotation, 
anticenter and $Z$ directions.}

After conversion to $v_{GSR}$, an even greater agreement is found (Fig. 17) between 
the actual velocities of the LMC
(big solid circle), the 332 km s$^{-1}$ group in the Carina field (smaller filled circles),  
and RGB stars we have found with similar velocities between these
systems (open triangles and circles).  
While the relative numbers of stars in each position on Figure 17
are a function of widely varying survey areas, spectroscopic magnitude limits, and 
spectroscopic target selection (i.e. whether or not stars were selected to be an LMC-like
giant, a Carina-like giant, or any kind of giant); what is relevant is the smooth variation
of the mean velocities in each survey field from the 
LMC to the 332 km s$^{-1}$ group, a trend that strongly suggests
a dynamical association of all of these stars.\footnote{An apparent difference in the
velocity dispersions among the different sets of points in Figure 17 is in part attributable
to the more than $5\times$ lower RV
precision of the measurements for the stars found outside the Carina survey region.} 
Even more intriguing is that this velocity trend matches that found for other LMC
tracers (e.g., \citealt{schommer1992}; \citealt{kunkel1997b})
at similar position angles from the LMC core over a 13$^{\circ}$ angular separation from the
LMC center, and where the trend is attributed to the rotation curve of the LMC 
(\citealt{schommer1992}; \citealt{kunkel1997b}; \citeauthor*{vdM02}).

Figure 17 shows the RV trend for the 
LMC disk (solid line) and halo (dashed line) from the best-fitted model to previously published 
outer LMC data by \citeauthor*{vdM02} 
(see their Fig. 5). 
We show the trends at LMC position angles corresponding to our 
survey fields and to the 13$^{\circ}$ limit of the 
model and previous data, as well as
an extrapolation of the \citeauthor*{vdM02} LMC RV trends to the Carina field.\footnote{The model in Figure 17
should not be interpreted as the actual rotation curve,
but a velocity trend on the sky.  The actual rotation curve
corresponding to these points is shown in Figure 6 of van der Marel et al. (2002).} 
This figure suggests that the inner data follow the disk velocity trend, whereas the 332 km s$^{-1}$ 
moving group lies right on the extrapolation of the halo velocity trend to $\sim22^{\circ}$ 
($\sim20$ kpc) radius from the LMC center.

To further test an association of the Magellanic Clouds to the 332 km s$^{-1}$ group stars, 
we compare in Figure 18 their distribution in the CMD and 2CD to that of
stars found in the closest survey field to the LMC, shown by a green open circle in
Figure 16.
The position of spectroscopically-confirmed
LMC stars from this same inner RV survey field are shown by green open triangles in Figure 18.
Figure 18a shows that the CMD position of the 332 km s$^{-1}$ group stars (red solid circles) 
is {\it precisely} where the locus of the LMC's prominent red clump 
slightly overlaps our Carina RGB selection boundary. 
Moreover, inspection of our Carina field sample of giant stars that fall outside our 
Carina CMD selection region in
Figure 14a reveals: (1) a possible additional concentration of stars at $M_{0} \sim 19$ 
just outside the 
Carina selection boundary at the position 
of the LMC red clump seen in Figure 18a (although not stretching as blueward in 
Fig. 14a because
such stars are eliminated by the 2CD selection); and (2) a slight excess of
stars tracking the nominal position of the LMC RGB visible in Figure 18a, above 
the Carina RGB selection
boundary.  To test whether both of these groups of ``Carina outliers" may be Magellanic in origin,
on UT 2005 August 15 
we observed two bright giant candidates in this 
``LMC RGB position" of the CMD (marked as solid squares in Figure 18)
using the MIKE spectrograph on the Magellan telescope.
These turned out to have RVs (317 and 342 km s$^{-1}$) consistent with
membership in the 332 km s$^{-1}$ group, which further
vindicates a Magellanic Cloud provenance of this moving group.\footnote {We note the RV uncertainties 
for these stars are large, $\sim15$ km s$^{-1}$, therefore we do not include them in
the velocity dispersion calculation but only use them as membership information.}
Comparison of Figures 14a and 18a 
certainly evokes the notion of
a diaphanous presence of LMC stars in the foreground of the Carina dSph, which
has given rise to the 332 km s$^{-1}$ group. 
Finally, within the GIRAFFE RV dataset, we found four more stars with velocities
matching the 332 km s$^{-1}$ group and positions in the CMD
(red open circles in Figure 18) reasonably compatible with being LMC 
red clump stars. 
Adding these four
stars changes only marginally the mean velocity and the velocity dispersion of the moving group. 

With the possible connection to the Magellanic Clouds 
in mind, we can bring the abundance argument full 
circle to look for self-consistency of this hypothesis.  For example,
if the originally identified 332 km s$^{-1}$ 
group members are parts of the red clump of the LMC, then for each star we can recalculate its [Fe/H] 
from the infrared triplet strength
assuming the $V_{HB} = 19.2$ of the LMC red clump.  The result 
yields a mean [Fe/H]=-0.67 (with 
dispersion $\pm0.62$ dex) --- relatively more metal poor than, but still consistent with
the mean metallicity 
([Fe/H]=$-0.37$) of the dominant population of stars in the LMC found recently by 
\citet{Cole05} using the same infrared triplet methodology.   
Considering also that it would seem unlikely to find two such extreme velocity 
stellar systems at a similar distance and position in the sky, the collective evidence
compellingly suggests that we have found widely dispersed stars from 
one of the Magellanic Clouds --- the LMC being more likely ---
in the foreground of the Carina dSph.

\subsection{Implications for LMC Structure}

As with the examples of the Carina dSph explored earlier, and the Ursa Minor
system explored in \citet{Mu05}, the presence of extremely widely displaced, but
satellite-associated stars would seem to have profound implications for the structure of the LMC.
One can use equation (45) from \citeauthor*{vdM02} to estimate the mass of the LMC given
a certain tidal radius.
For our most widely separated star in the 332 km s$^{-1}$ group ($\sim22^{\circ}$
away from the center of the LMC)
to remain bound
to the LMC implies a minimum LMC mass of 
$3.1\times10^{10}$ M$_{\sun}$ assuming a Milky Way mass interior to the LMC of
$4.9\times10^{11}$ M$_{\sun}$  (\citealt{kochanek96}; the \citealt{Burkert1997} model 
gives almost the identical Milky Way mass). This inferred LMC mass is
$\sim3.5$ times more than that reported by
\citeauthor*{vdM02} ($8.7\times10^9$ M$_{\sun}$) to a 13$^{\circ}$ radius
and consistent with the $2.0\times10^{10}$ M$_{\sun}$ LMC mass derived
if we assume a flat LMC rotation curve to this distance.  
The implied 20.2 kpc minimum
tidal radius is now more than 33\% (1$\sigma$) greater than the $15.0\pm4.5$ kpc tidal
radius estimated by \citeauthor*{vdM02}.

These results immediately suggest
two possible scenarios (ignoring possible 
solutions offered by Modified Newtonian Dynamics; \citealt{Milgrom1995}; \citealt{sanders2002}): 
(1) The LMC is substantially larger
than previously appreciated.  The inferred total $M/L$ would exceed
10 in solar units.
An even larger mass is implied by the fact that the velocity dispersion of the 332 km s$^{-1}$
--- 9.8 km s$^{-1}$ --- while $\sim 2\times$ smaller than the dispersions of tracers 
$< 10$ kpc from the LMC, as might be expected in the outer limits of a galaxy halo, are
still quite larger than the expected, small asymptotic value at the ``edge" of a 
galaxy.\footnote{While RVs for stars in our analysis that lie
outside the Carina field are generally of lower resolution,
the velocity dispersions for our fields less than 10$^{\circ}$
match well those found for the carbon stars summarized in Figure 6 of \citeauthor*{vdM02}.}
We note that an LMC extending out to $\sim20$ kpc (in the line of sight) has been already 
proposed by \citet{Zaritsky1997} based on the identification of a vertically extended
red clump in the CMD of a field in the direction of the LMC.
(2)  The Magellanic Cloud stars we observe in the foreground are not bound
to the LMC.  The colder dynamics of the 332 km s$^{-1}$ stars might 
be explained through a tidal debris origin. 

But if unbound, stars in the direction of the Carina dSph are {\it not} aligned
with the expected direction of an LMC tidal tail, based on both the typical 
proper motions\footnote{We must note that this may not be a problem if the
LMC had a significantly different proper motion. In particular, \citealt{momany2005}
argue that the LMC is in fact moving in the direction of Carina, but warn the reader
that there are likely to be unidentified systematic errors in the UCAC2 that they 
used that are responsible for these results.}
measured for the LMC (summarized in Table 1 of \citeauthor*{vdM02}) 
as well as the direction of the HI Magellanic Stream (both the leading and trailing arms) --- both lie 
in a roughly orthogonal direction.
This is not necessarily a problem, since stars will be tidally stripped 
anywhere along the satellite-Milky Way equipotential, whereas we have only explored one position angle
from the LMC here. 
On the other hand, the Carina survey field {\it does} happen to lie more or less along the axis 
defined by the LMC and SMC.  A tidal disruption
scenario involving an interaction of the LMC and SMC might conceiveably throw Magellanic stars
out along this axis.  For example, the velocities of our Magellanic giant stars are consistent with those
of the carbon stars found by \citet{kunkel1997a,kunkel1997b}
in the same general direction (see Figs. 16 and 17), and which these
authors attribute to a ``polar ring" of SMC debris around the LMC.
Alternatively, the widely separated ``LMC" stars may constitute
residue from the disruption of a former ``Greater Magellanic Galaxy" which has often been
invoked as a possible explanation for the curious alignment of a number of Milky Way
satellites and globular clusters along a ``Magellanic Plane" that also includes the
HI Magellanic Stream
(\citealt{kunkel79}, \citealt{lyndel82}, \citealt{maj94}, \citealt{fusi95},
\citealt{LandL95}, \citealt{MPR96}, \citealt{palma2002}).  A dynamical association 
of Ursa Minor, Draco, the LMC and the SMC is suggested by their common motions along one great circle  
(see, e.g., Fig. 3 of \citealt{palma2002}).
Were this group of Milky Way satellites
truly daughters of the break up of a Greater Magellanic system or produced together as tidal 
dwarfs during a major merger with the Milky Way,
their close, but not precise,
alignment in a single plane might indicate the possibility of
a potentially broad stellar swath of loosely coherent Magellanic Plane debris.  
But if the
332 km s$^{-1}$ stars represent dynamically {\it old} tidal debris like this, one might
not expect it to so well match the current distance of the LMC, nor its velocity
(or, even more coincidentally, the velocity extrapolated from the LMC velocity trend 
to this position in the sky).

Only with further surveying for additional
``332 km s$^{-1}$ group" stars in other directions around the Magellanic Clouds
can one hope to test such hypotheses.  
We intend to explore these possibilities further elsewhere (Nidever et al., in preparation)
with a larger database of outer LMC stars collected over a larger area.

\section{Summary and Discussion}

Our survey for diffuse halo substructure in a large field around the 
Carina dSph has yielded the following primary results on the 
structure of both the Carina dSph and the LMC (or Magellanic Clouds):  

--- Using a combination of new Washington+$DDO51$ photometry and new echelle spectroscopy
we have confirmed the existence of an extended, power law component in the density
distribution of Carina, which can be modeled as a ``King + power law".  Such density
distributions are characteristic of those found in models of disrupted satellites and has
also been observed in the tidally disrupting Sgr dSph.

--- With Magellan+MIKE echelle spectroscopy of giant
star candidates in the Carina field we have establish the existence of Carina
stars to the limits of our photometric survey field, with confirmed Carina members to
at least 4.5$r_{lim}$, and likely 4.9$r_{lim}$.  These detections represent
the most widely separated stars (in terms of $r_{lim}$) 
found associated with any dSph (apart from the Sgr dSph) to date.  
Beyond verifying the existence of the extended Carina population, these 
widely separated member stars 
have profound implications for the structure of Carina:  If the stars are bound, Carina must
have a minimum total $M/L$ of 6,300 in solar units, or 16,000 in the case of 
the 4.9$r_{lim}$ example.  

---  With the addition of VLT+GIRAFFE spectroscopic data and other published data (\citealt{Mateo1993}; \citeauthor*{Paper VI}) within $r_{lim}$ to our MIKE velocities at large $r_{lim}$, 
we have good and continuous sampling of the Carina RV distribution to well past $r_{lim}$ by 408 
confirmed Carina members.  With these data,
we have rederived the central and global $M/L$ for Carina, assuming a single-component model
and using the core-fitting technique; the results yield 43$^{+53}_{-19}$ and 41$^{+40}_{-25}$ 
(M/L)$_{\sun}$
respectively, where the main source of uncertainty comes from the luminosity.
These results are significantly at odds with the lower limits to the global $M/L$ found using
the outlying Carina members above.

--- With the extensive RV coverage
we have also derived the line of sight radial velocity dispersion profile for Carina
to $\sim2.5r_{lim}$, the most extensive such profile so far (by more than a factor
of two) for any dSph.
The profile is flat to past $r_{lim}$ and then exhibits a rise in the dispersion
to almost twice the inner value at $> 2r_{lim}$. 
Such results are incompatible with completely bound, mass-follows-light dSph models, but also
challenge two-component
models that account for the flat dispersion via an extended dark halo
surrounding the dSph.  In the latter case an enormous halo is needed, one 
significantly more massive than that implied above for simply keeping the $>4 r_{lim}$
Carina stars bound, since
the significant dispersion at large radius implies that the tidal radius is much farther out.

--- While with our new data 
we cannot definitively rule out a {\it very} large, and extended dark halo for Carina --- one producing
a global $M/L$ approaching as much as 6,300 or more --- we conclude that a simpler, 
less contrived scenario that provides a good match to {\it all} available observations of Carina is that 
it is tidally disrupting and we have identified some of its unbound stars.  This scenario simultaneously
accounts for the following observed features of the Carina system: 
(1) Its ``King+power law" density profile, which is a natural product of tidal disruption;
(2) the fact that
the extended component of Carina lies predominantly along its major axis and shows increasing
ellipticity with radius, as would be expected in nascent tidal tails; 
(3) Carina stars extending from the core to the edge of the survey area;
(4) the flat, then rising velocity dispersion profile with radius; 
and (5)
a flattening of the RV distribution with radius, from Gaussian in the core to platykurtic at large
radius.  Explaining this combination of observed Carina properties with
extended dark halo scenarios will require substantial efforts to create successful ad hoc models.
On the other hand, all of the above Carina features not only 
resemble those seen in the established, tidally
disrupting Sgr dSph system, but have been well-matched by mass-follows-light models
(presented in a companion paper, Mu\~noz et al. 2006)
of disrupting dSphs having the nominal central $M/L \sim 40$ derived here.

--- Finally, we have detected a second, strongly velocity-coherent
structure in the Carina field with even higher RV than Carina.
The more metal rich stars constituting this moving group have CMD positions 
consistent with LMC red clump stars and their velocities follow the extrapolated 
velocity trend expected for LMC halo stars.  With additional Washington$+DDO51$
photometry and follow-up spectroscopy we have traced this population from 4$^{\circ}$
separation from the center of the LMC out to the 22$^{\circ}$ separation of the Carina field center.
These stars either represent the detection of Magellanic stellar tidal debris, or, if bound
to the LMC, imply a significantly larger mass and tidal radius for the LMC than previously determined.

Traditionally, debate over the  
kinematical and structural properties
of the diffuse, low surface brightness dSphs has tended to polarize around two primary 
interpretations: (1) that dSphs 
are dark matter dominated (e.g., \citealt{Mateo1998}) 
galaxies, with $M/L$ reaching to as much as 100 (M/L)$_{\sun}$, making them structurally different 
compared to globular clusters and dE systems.
The prime observational evidence to support 
this claim is the relatively high measured central velocity dispersions
that --- coupled with an assumption of dynamical equilibrium ---
imply masses far in excess of that inferred by the luminous component.  Alternatively, (2)
dSphs have also been discussed as systems partly or completely out of 
virial equilibrium 
(\citealt{hm69}; \citealt{kuhn1989}; \citealt{kuhn1993}; \citealt{Kroupa1997};
\citealt*{GFM99}; \citealt{fleck2003}). 
Such an assertion seeks to explain the large central velocity dispersions of dSphs through
inflation by tidal heating or other dynamical effects, 
allowing for much more modest dSph masses, consistent with no dark matter.

To date, despite much observational and theoretical effort,
the physical evidence has generally remained unpersuasive enough to
dislodge the most ardent adherents to these models.
Reinforcing viewpoints have been several ``all or nothing" notions
introduced into the debate, including: (1) the assumption that dark matter dominated systems
are in dynamical equilibrium throughout their entire physical extent (e.g., \citealt{Stoehr2002};
\citealt{Walker2005});
or, (2) if evidence of tidal stripping is found around a dSph, the system must be devoid of 
dark matter (e.g., \citealt{Burkert1997}).

Remarkably, more recent work intended to {\it clarify} the physical nature of dSphs
has, instead, increased the apparent gulf between diametrical viewpoints.
\citet{Kleyna1999} had previously suggested that ``only $\sim10-20$ additional
observations [of dSph star RVs] at 0.75 times the tidal radius would be required to 
distinguish clearly between an MFL distribution and an extended halo or disrupted
remnant model with a flat or radially rising velocity dispersion."
Yet, despite the fact that the latest dSphs
spectroscopic surveys have provided RVs of hundreds of dwarf members to beyond
0.75$r_{lim}$ in several systems
(\citealt{Mateo1997}; \citealt{Kleyna2002,Kleyna2004}; \citealt{W04};
\citealt{Tolstoy2004}; \citealt{Westfall2006}; \citealt{Mu05}; \citealt{Walker2005}; 
\citealt{Sohn2006}), we are apparently no closer to a consensus view of dSph dynamics. 
While this is partly due to technical differences in interpretation of even the same
databases (e.g., Wilkinson et al. 2004, {\L}okas et al. 2005, Mu\~noz et al. 2005), in 
general most studies are finding 
flat dSph velocity dispersion profiles to be the norm.  As discussed several times here, such profiles
are produced naturally in dSph models  
undergoing tidal disruption (see also \citealt{kuhn1989}; \citealt{Kroupa1997};
\citealt{Mayer2002}, \citealt{fleck2003}; \citealt{Read2005a}; M06).
However, rather than settling the issue, 
these rather flat velocity dispersion profiles have prompted the development
of even more extreme, two-component, extended dark halo dSph models 
with {\it substantially higher} bound masses and total $M/L$  --- exceeding 400 or even 1000 
$(M/L)_{\sun}$
(e.g., \citealt{Lokas2002}; \citealt{Kleyna2002}; \citealt{Walker2005}). 
These models are partly motivated by the proposition
that the so-called missing satellite problem of $\Lambda$CDM cosmologies
could be alleviated if galactic dSph satellites inhabit the most
massive sub-halos --- i.e., M$_{dSph}>10^{9}$ M$_{\sun}$, or 
equivalently V$_{circ} >$ 30 -- 40 km s$^{-1}$
(\citealt{Stoehr2002}; \citealt{Hayashi2003}).\footnote{We note, however, that
\citet{Kazantzidis2004} argue against this picture, showing that, in the case of 
Draco and Fornax, only halos of V$_{circ} <$ 25 km s$^{-1}$ can succesfully reproduce the 
velocity dispersion profiles of these dSphs.}

From the numerous arguments laid out thus far, we are persuaded that the weight of evidence
militates against the extreme halo hypothesis for Carina in favor of a tidal disruption scenario.  
Yet one more argument favors the latter hypothesis:
An extended DM halo of the magnitude our data would require in this scenario
has ancillary implications for the {\it chemical evolution} of Carina that are problematical.
Despite having a complex and episodic star formation history, Carina
has a relatively low mean metallicity of [Fe/H]$\sim-1.9$ (\citealt{monelli2003}; \citealt{Koch06}).
\citet{Tolstoy2003} note that galaxy masses of order a few times $10^{7}$ M$_{\sun}$ 
--- consistent with the mass of Carina derived from
central velocity dispesion (\citealt{Mateo1993}; \S 4.4)  --- are low enough
to suffer metal-enriched winds, which promote preferential depletion of metals but
retention of sufficient gas to allow further star formation at a continued, relatively low mean metallicity
(like Carina's).  
A larger galactic potential diminishes the possibility of blow-out/blow-away of either gas
or metals (e.g., \citealt{Vader1986,Vader1987}; \citealt{MLF1999}; \citealt{Ferrara2000}),
leading to closed-box enrichment
(\citealt{Tolstoy2003}).  
But \citet{Koch06} find that the metallicity distribution
function of Carina is not well matched by a closed-box model.
If Carina has an enormous extended DM halo, it
would have resulted in an enrichment that it is not observed (\citealt{S-H1996}).
However, a more modest Carina dark matter content is not discounted by this argument.

One of our goals in this paper (see also \citealt{Mu05}) has been
to push the measurement of physical parameters in one dSph
to hitherto unexplored regions to see if, in at least one
system, new data in the extrema can definitively rein in the range of possible 
models.  We conclude that the new breed of extremely large $M/L$, extended dark matter
halos is less likely to apply to the present Carina dSph than a tidal disruption scenario, 
which more readily explains all present observational data on the satellite.\footnote{This conclusion does
not preclude the possibility that a formerly extended dark halo might have been stripped from Carina
at earlier times.  Thus, the success of tidally disrupting, 
mass-follows-light models in describing at least some dSphs (M06, \citealt{Sohn2006})
could be consistent with $\Lambda$CDM if the models 
produce subhalos that are sufficiently stripped to reach the luminous matter.} 
That said, our results do not rule out {\it any} dark matter in the dSph, and, indeed,
as we shall show in Mu\~noz et al. (2006), an easily workable (and therefore likely) model
for Carina is one with 
elements of {\it both} of the originally debated dSph scenarios:
a tidally disrupting, mass-follows-light dSph, but one with relatively high ($M/L \sim 35-40$)
dark matter content, as suggested by the central velocity dispersion.

It has recently been claimed (\citealt{Gilmore2004}) that ``Sgr was a rare event, 
not a paradigm for the average''.  This conclusion has been motivated by the notion that
were systems like the Carina dSph tidally disrupting, then the halo should have large numbers of 
youngish (e.g., blue main sequence) stars in larger numbers than seen 
(\citealt{UWG1996}).  Such an analysis presumes that the {\it present} Carina system is
representative of the typical stellar population that would have been contributed to the
halo by tidally disrupting dSph systems {\it including the former Carina}.   
In contrast, as was
previously demonstrated in \citet{Majewski2002}, if 
for a Hubble time 
Carina were disrupting at the fractional mass loss rate implied by its density profile,
--- i.e. $<0.24$ Gyr$^{-1}$ (see \S 2.2) ---  
then far more stars from Carina's oldest population
would have been lost by now than 
from either the intermediate-aged or young populations in Carina.  This is also a reasonable
explanation for why the Carina system today is dominated by it's intermediate-aged 
population, and even for why there seems to be a radial metallicity and age
gradient in Carina: 
The present balance of populations bound to Carina likely reflects the
competing interplay of star formation history and mass loss history (\citealt{Font2006})
in this disrupting analogue of the Sgr dSph galaxy.

We appreciate useful discussions with David Law and Andrew
McWilliam.  We thank an anonymous referee  for
helpful suggestions that helped improve the paper and to Carlton Pryor
for providing us with his code to calculate velocity dispersions
using the Maximim Likelihood Method.
We acknowledge funding by NSF grant AST-0307851, NASA/JPL contract 1228235, 
the David and Lucile Packard Foundation, and the generous support of Frank Levinson
and through the Celerity Foundation.  
Additionally, P.~M.~F. is supported by the
NASA Graduate Student Researchers Program, a University of Virginia
Faculty Senate Dissertation-Year Fellowship,
and by the Virginia Space Grant Consortium. D.L.N. is supported by the ARCS 
Foundation, a University of Virginia President's Fellowship, and by the Virginia 
Space Grant Consortium.

\clearpage
\begin{deluxetable}{ l c c c c c c c c c c c } 
\tabletypesize{\scriptsize}
\tablewidth{0pt}
\tablecaption{Radial Velocities of Stars observed with MIKE\tablenotemark{a}}
\tablehead{ \colhead{Star} &
\colhead{$\alpha_{2000}$} &
\colhead{$\delta_{2000}$} &
\colhead{UT Date } &
\colhead{$M_o$ } &
\colhead{$(M-T_2)_o$ } &
\colhead{$(M-DDO51)_o$ } &
\colhead{$RV_{helio}$ } &
\colhead{$RV_{gsr}$ } &
\colhead{Q/$\sigma$} &
\colhead{Cand?\tablenotemark{b}} &
\colhead{Member}}
\startdata
 C830356 &   6:35:11.95 &   $-$51:25:05.1 & 27Jan2004 &  18.84 &   1.30 &  $-$0.00 &  328.0 &  119.7 &   6 & Y & N \\
C2520066 &   6:36:40.86 &   $-$51:58:07.0 & 27Jan2004 &  19.04 &   1.39 &   0.02 &  213.9 &    5.3 &   7 & Y & Y \\
C2640634 &   6:38:22.77 &   $-$51:11:00.4 & 27Jan2004 &  18.08 &   1.62 &   0.01 &  221.2 &   12.2 &   7 & Y & Y \\
C2680057 &   6:38:36.82 &   $-$51:16:23.9 & 28Jan2004 &  18.33 &   1.57 &   0.03 &  222.1 &   13.1 &   2.6 & Y & Y \\
C2411078 &   6:38:47.04 &   $-$50:50:31.2 & 28Jan2004 &  18.65 &   1.42 &   0.04 &  229.4 &   20.4 &   4 & Y & Y \\
\enddata
\tablenotetext{a}{A full version of this table can be found in the electronic edition of the 
{\it Astrophysical Journal}.}
\tablenotetext{b}{Denotes a photometric Carina giant candidate.}
\end{deluxetable}
\clearpage

\begin{deluxetable}{ l c c c c c c c c c c c }
\tabletypesize{\scriptsize}
\tablewidth{0pt}
\tablecaption{Radial Velocities of Stars observed with VLT/FLAMES\tablenotemark{a}}
\tablehead{ \colhead{Star} &
\colhead{$\alpha_{2000}$} &
\colhead{$\delta_{2000}$} &
\colhead{$M_o$ } &
\colhead{$(M-T_2)_o$ } &
\colhead{$(M-DDO51)_o$ } &
\colhead{$RV_{helio}$ } &
\colhead{$RV_{gsr}$ } &
\colhead{$\sigma$} &
\colhead{Cand?} &
\colhead{Member}}
\startdata
C2501778 &   6:38:31.92 &   $-$51:07:04.8 &   19.76 &   1.22 &   0.01 &  230.3 &   21.3 &   2.9 &  2CD\tablenotemark{b} &    Y \\
C2680118 &   6:38:45.15 &   $-$51:11:26.6 &   19.56 &   1.20 &   0.02 &  216.3 &    7.2 &   2.6 &    Y &    Y \\
C1400762 &   6:38:47.04 &   $-$51:00:46.8 &   18.60 &   1.27 &  $-$0.07 &   20.8 & $-$188.2 &   1.1 & CMD2\tablenotemark{c} &    N \\
C1401432 &   6:38:54.60 &   $-$51:04:01.2 &   20.17 &   1.19 &   0.05 &  216.8 &    7.7 &   7.6 &    Y &    Y \\
C1402042 &   6:39:03.60 &   $-$50:57:43.2 &   20.63 &   1.91 &  $-$0.02 &   13.8 & $-$195.3 &   2.1 & CMD1\tablenotemark{d} &    N \\
C2413901 &   6:39:11.88 &   $-$50:58:40.8 &   19.28 &   1.31 &  $-$0.11 &   69.5 & $-$139.7 &   1.0 &    N &    N \\
C2413890 &   6:39:12.60 &   $-$50:54:10.8 &   19.68 &   1.19 &   0.02 &  225.0 &   15.8 &   3.4 &    Y &    Y \\
\enddata
\tablenotetext{a}{A full version of this table can be found in the electronic edition of the
{\it Astrophysical Journal}.}
\tablenotetext{b}{Denotes a star that is a 2CD outlier but lies inside Paper I giant box and within the
Carina RGB selection.}
\tablenotetext{c}{Denotes a star that lies just outside the RGB selection, but is a giant stars according
to the Paper I selection criteria.}
\tablenotetext{d}{Denotes a star that lies just outside the RGB selection, but is a giant stars according
to the most conservative giant box used in this paper.}
\end{deluxetable}
\clearpage

\begin{figure}
\plotone{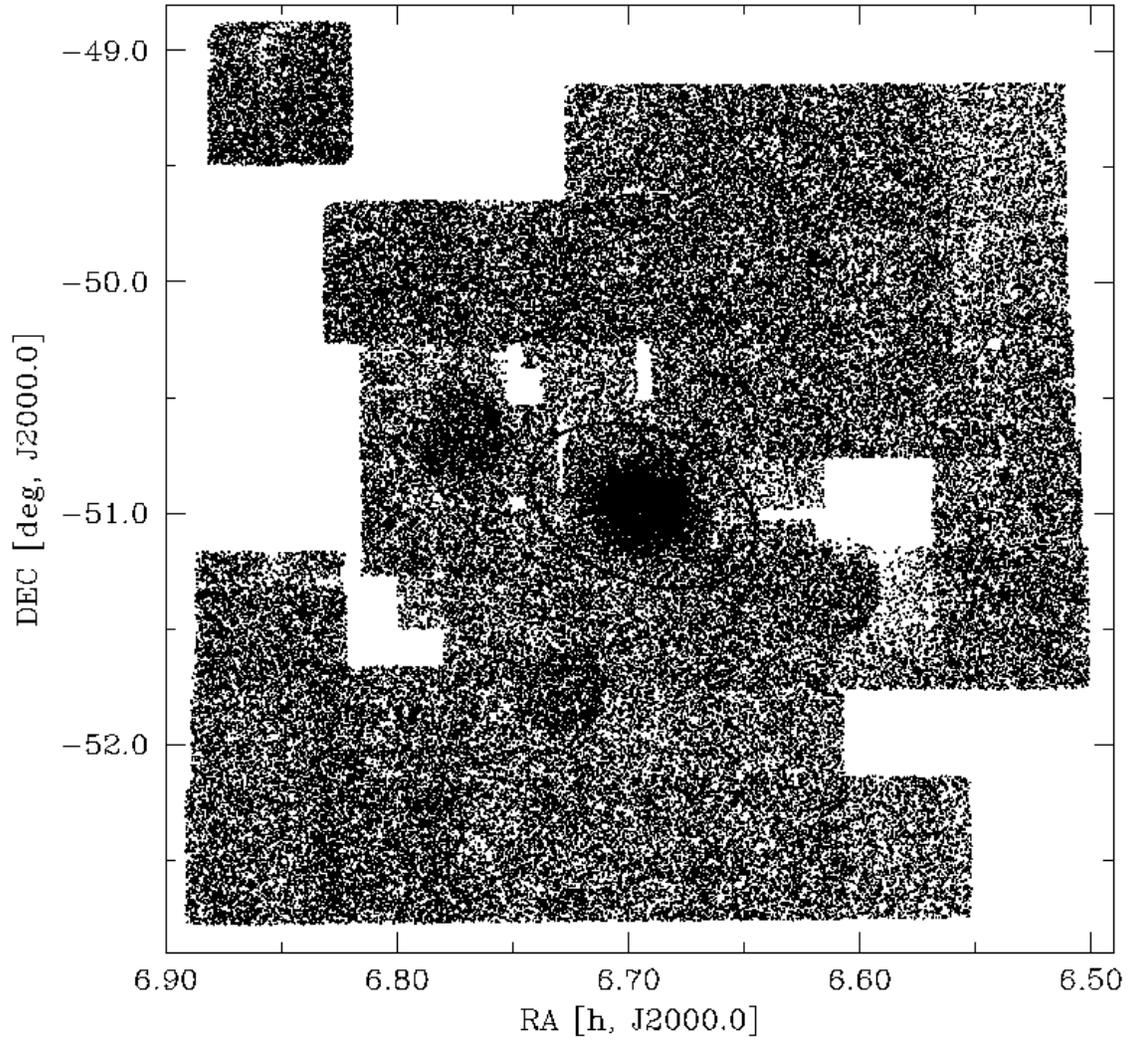}
\caption{The area covered by our new photometric and spectroscopic survey. Points represent 
all stars in our photometric survey brighter than $M > 21$.  
The nominal King limiting radius of Carina from \citeauthor*{IH95} is delineated
by the ellipse.}
\end{figure}

\begin{figure}
\plotone{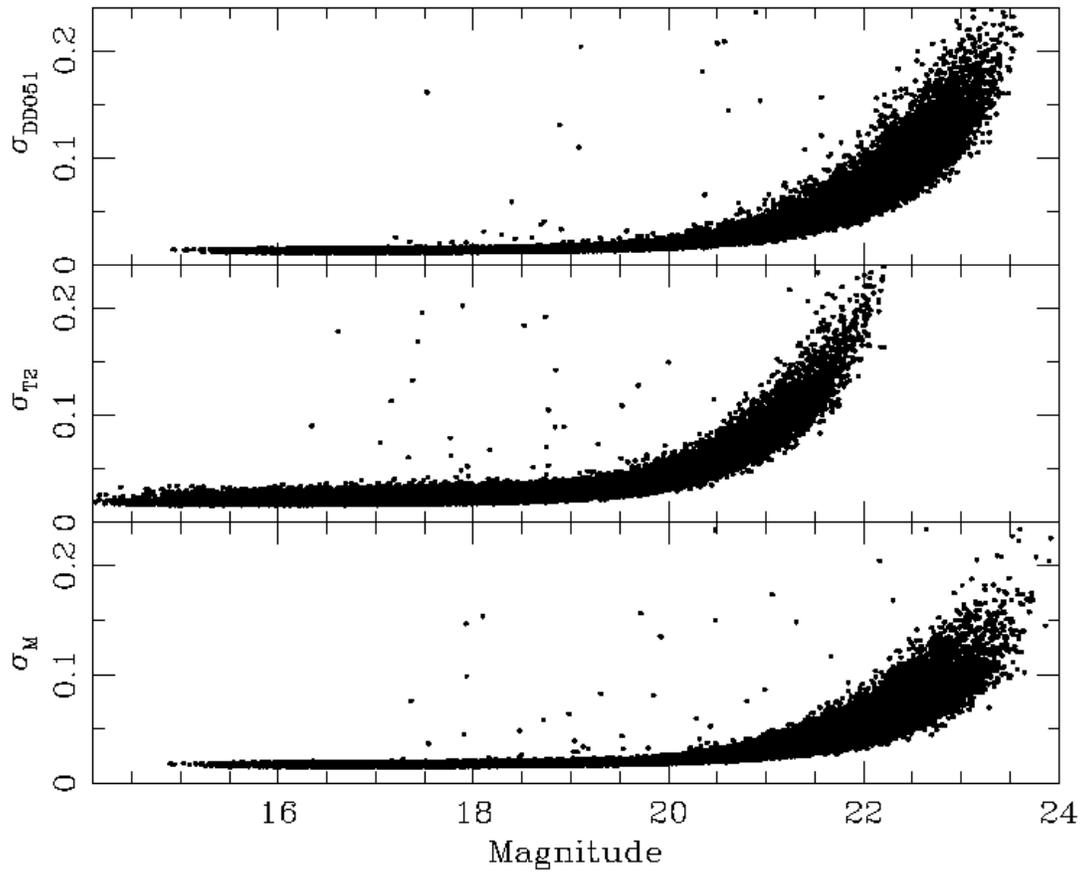}
\caption{Photometric errors as a function of magnitude in our new Mosaic camera survey of the field
centered on the Carina dSph.}
\end{figure}

\begin{figure}
\plotone{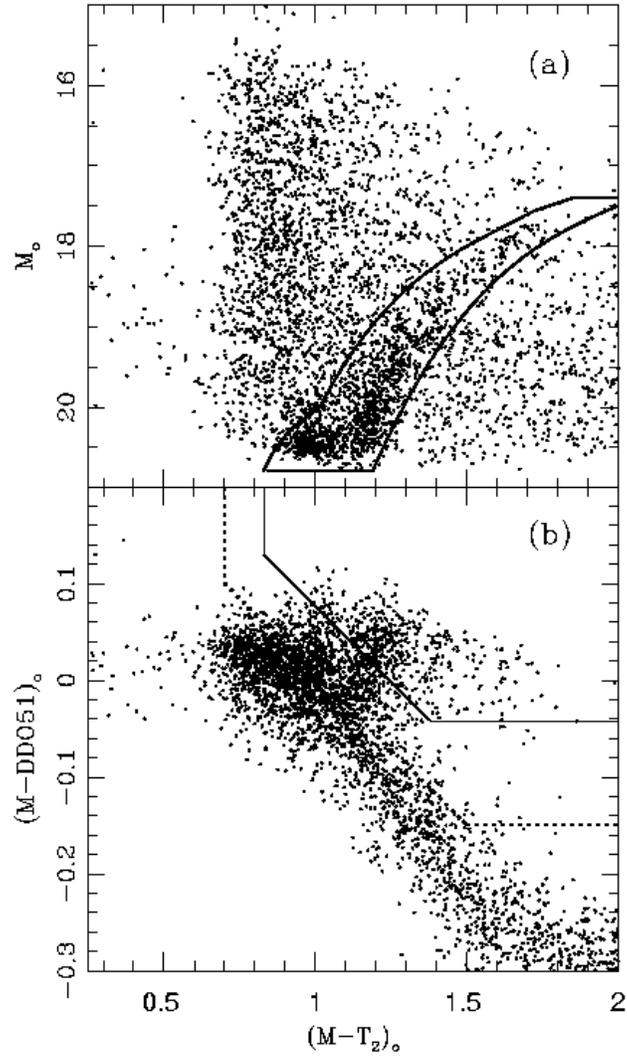}
\caption{(a) The color-magnitude diagram of stars in our new Carina photometric survey.  The solid line
marks the region adopted to represent the Carina red giant  branch.
(b) The two-color diagram of stars in our Carina survey to $M=20.8$.
The solid line marks the region from which we pick stars likely to be giant stars.
The dotted line delineates an expanded selection criterion used in \citeauthor*{Paper I} and 
explored in \S 2.2 and \S 4.1. In both panels, only
stars within one King limiting radius have been plotted as a guide to the
general features of these distributions.}
\end{figure}

\begin{figure}
\plotone{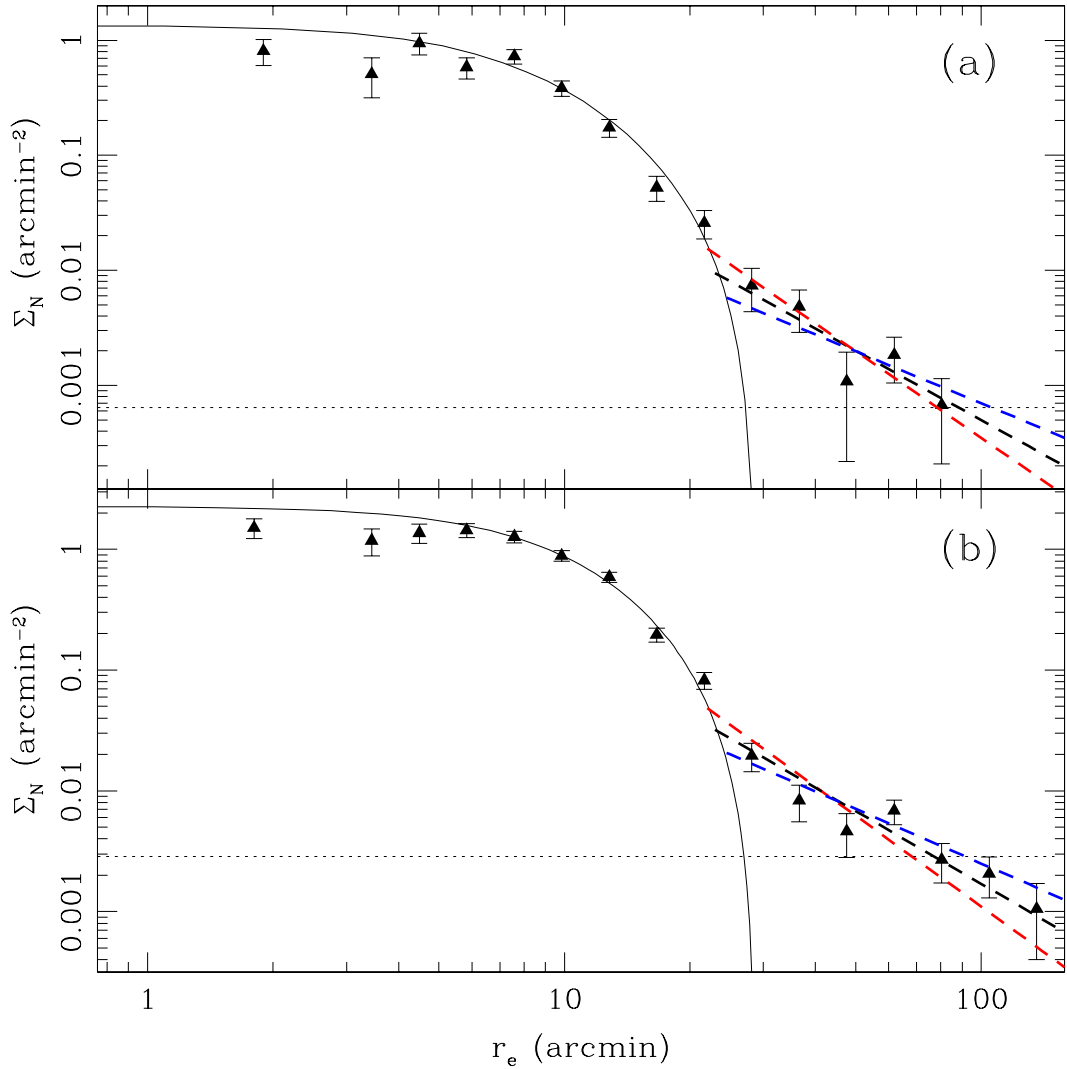}
\caption{(a) The density profile derived for the Carina dSph using the $T_2 \le 18.4$ sample.
The dotted line shows the equivalent background density level that has been subtracted to produce
the profile, as measured directly from the spectroscopic sample outside the nomimal
King limiting radius (the ellipse shown in Figure 4, adopted from \citeauthor*{IH95}).  
All stars have
been binned into elliptical annuli according to the Carina ellipticity and position angle 
derived by \citeauthor*{IH95}.  Outside the \citeauthor*{IH95} King 
limiting radius, the density values shown 
come almost directly from the spectroscopically-confirmed Carina members from \S3,
which is 90\% complete to $T_2 = 18.4$.
(b) Same as panel (a) but for the $M \le 20.8$ giant candidate sample.  In this case, 
where we do not have spectroscopic coverage to the magnitude limit, 
we have subtracted the mean background level as derived by the method described in the
text.
The dashed lines in both panels show $r^{-1.5}$, $r^{-2}$ and 
$r^{-2.5}$ power laws, while the curving solid 
line shows the \citeauthor*{IH95} King profile, 
scaled vertically to the density of our point at a radius of 10 arcmin.}
\end{figure}

\begin{figure}
\plotone{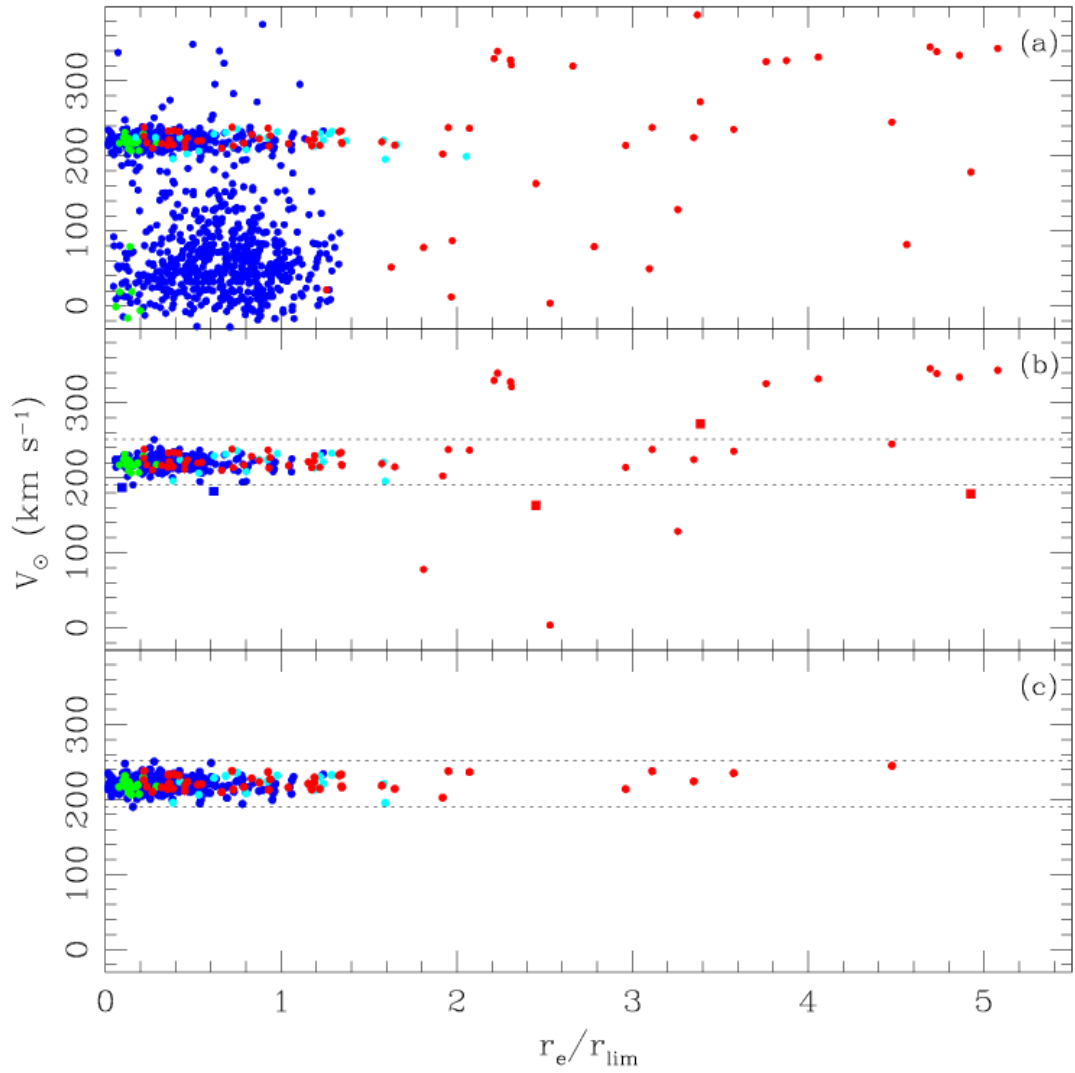}
\caption{Distribution of derived RVs for the newly observed Carina giant candidates as
a function of elliptical distance from the center of Carina.
(a) All stars observed with MIKE (red), VLT+GIRAFFE (blue), \citealt{Mateo1993} (green) and Carina
giants from \citeauthor*{Paper VI} (cyan). (b) Stars 
selected to be Carina giant candidates by our Washington+DDO5 photometry method and 
our CMD selection. The dotted horizontal lines mark the 3$\sigma$
boundary used as our RV selection criterion.
(c) RV distribution of our adopted final sample.}
\end{figure}

\begin{figure}
\plotone{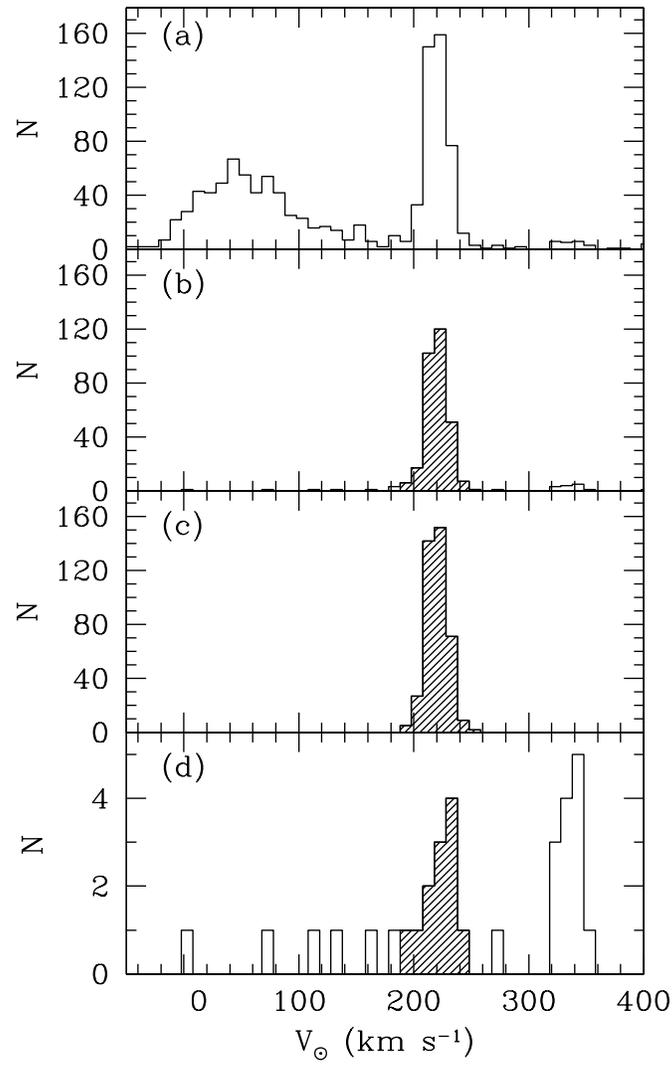}
\caption{Histogram of the radial velocities for (a) all stars. This includes stars observed 
with MIKE, VLT+GIRAFFE, stars from \citet{Mateo1993} and Carina giants from 
\citeauthor*{Paper VI}.
(b) Stars that have been selected to be Carina giants by our photometric method.
(c) Our final sample of Carina stars. (d) Histogram of stars selected as giant candidates 
having $r_{e} > 1.5r_{lim}$
We have shaded the region whithin our RV selection criterion as shown in Figure 5.}
\end{figure}

\begin{figure}
\epsscale{.9}\plotone{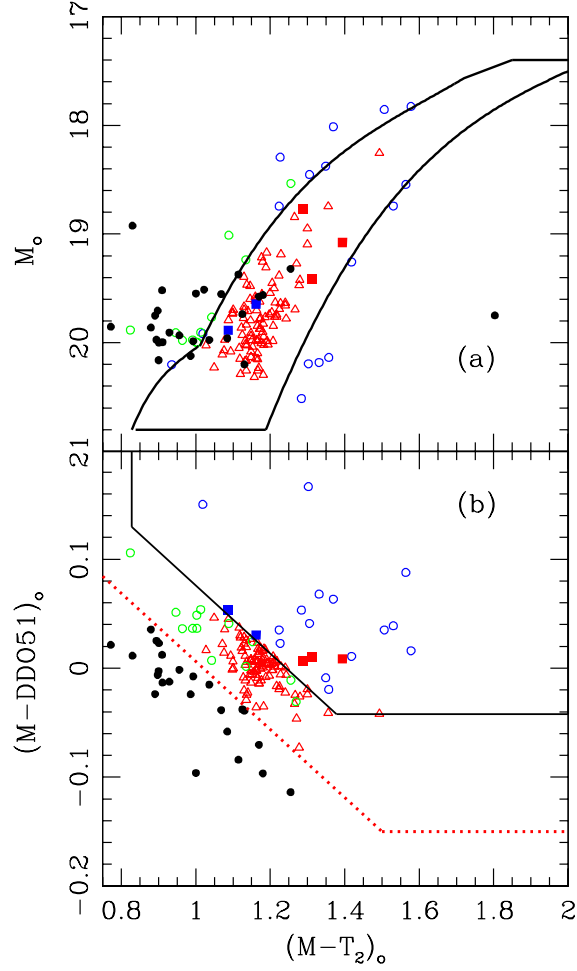}\epsscale{1}
\caption{(a) Color Magnitude Diagram and (b) Color-Color Diagram for VLT+GIRAFFE stars
with Carina-like velocities (as defined in Figure 5) that are not classified as
Carina giants.
Blue open circles mark stars that are CMD outliers, but that are classified as
giant candidates by the more conservative criterion shown by solid lines in panel (b).
Red open triangles show stars within the Carina RGB that were not selected as
giant candidates by our conservative 2CD selection, but that are within the
giant selection box defined in \citeauthor*{Paper I}.
Green open circles mark CMD and 2CD outliers by the conservative criteria of this paper but
that would have been classified
as giant candidates by the 2CD selection criterion adopted in \citeauthor*{Paper I}.
In addition, black circles show the CMD and 2CD position for stars with 
Carina-like velocities but that were not classified as giant by any of the above criteria.
The figure also includes (as solid squares) the CMD and 2CD positions for the 
RV outliers discussed in \S4.1.2 (blue squares for VLT+GIRAFFE examples 
and red squares representin examples from the MIKE data).
See the text for more discussion.}
\end{figure}

\begin{figure}
\plotone{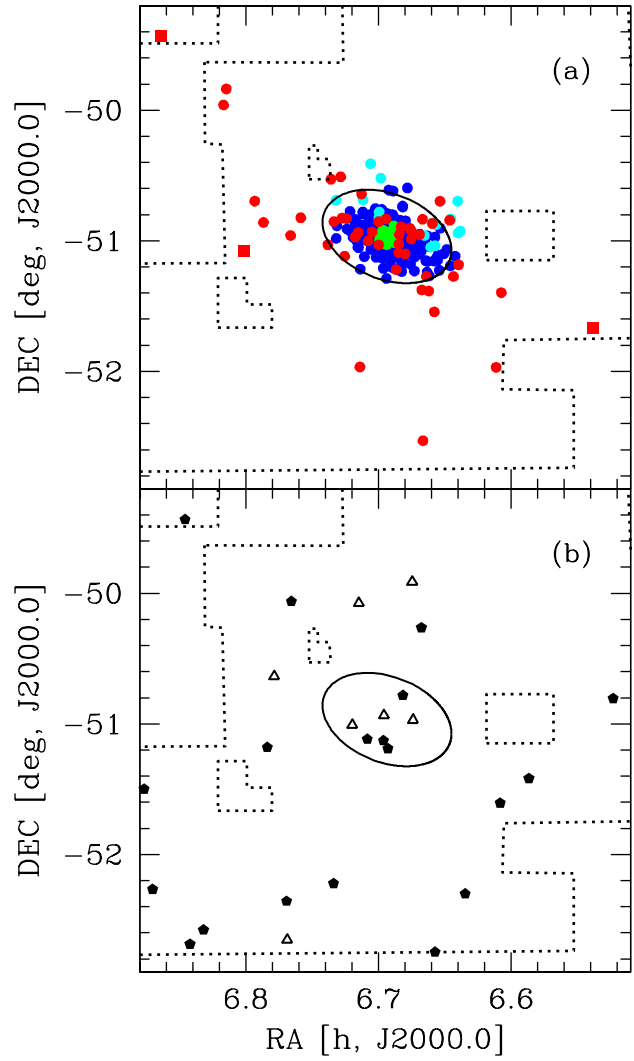}
\caption{(a) Distribution of bona fide Carina RV-members on the sky.
As in Figure 5, 
red circles show MIKE stars, blue circles mark 
VLT+GIRAFFE stars, green circles are stars from
\citet{Mateo1993} and cyan circles mark stars whose RVs come from \citeauthor*{Paper VI}.
Shown also in the figure are the three red squared from Figure 5, which
denote the three MIKE RV outliers discussed in \S4.1.3.  (b) The distribution of 
stars selected to be Carina 
giant candidates based on the CMD and 2CD that do not have Carina-like RVs is
also shown. In particular, solid symbols mark the distribution of 332 km s$^{-1}$ 
stars (see \S 6), and open symbols other RV outliers.} 
\end{figure}

\begin{figure}
\plotone{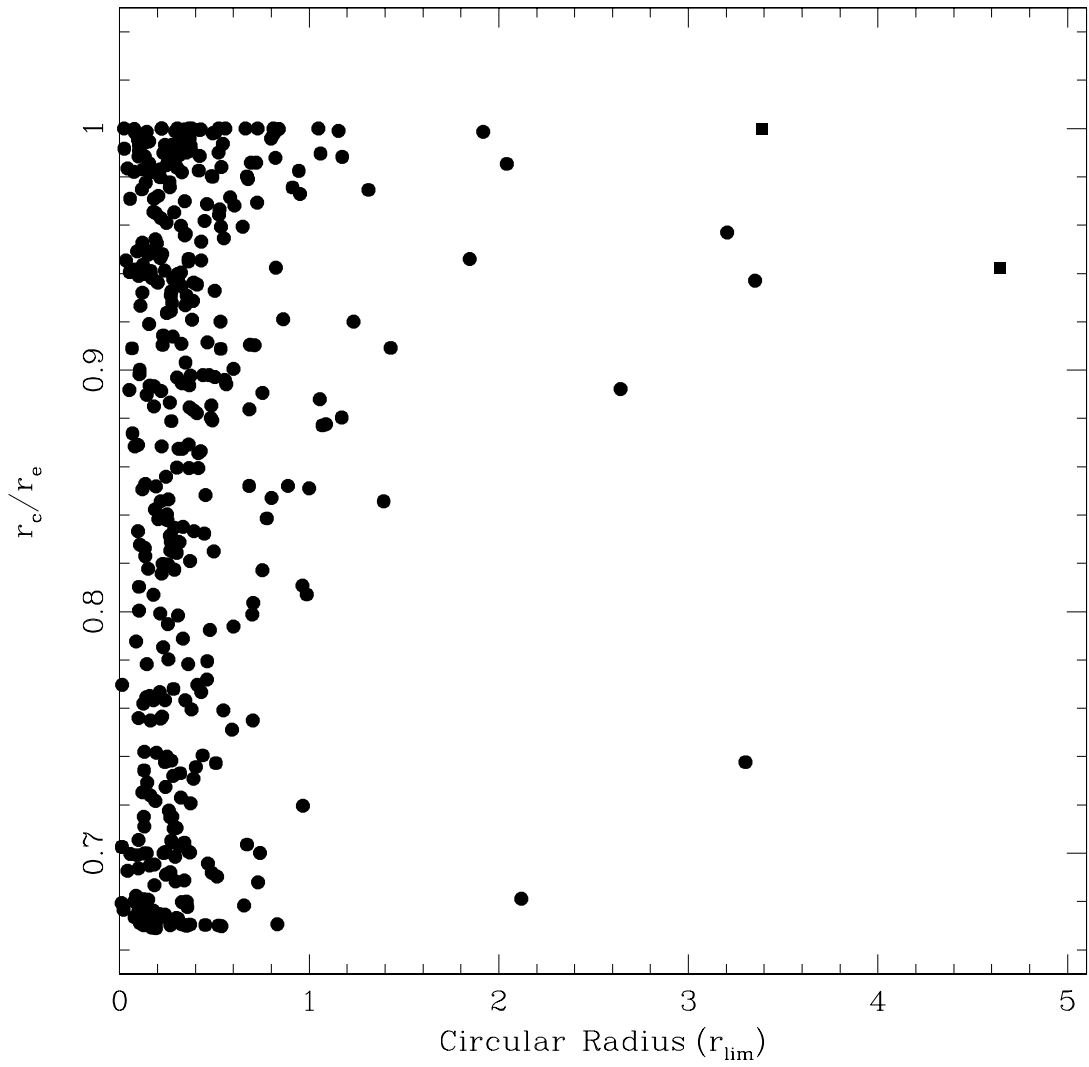}
\caption{Ratio of the circular to elliptical radius for each Carina
RV member versus circular radius. Stars on the major axis will have $r_{c}/r_{e}$=1
whereas stars on the minor axis will have $r_{c}/r_{e}$=0.67
(the ellipticity of Carina). The Figure shows that the mean
$r_{c}/r_{e}$ increases outward, indicating a preference of the
stars to lie along the major axis.}
\end{figure}

\begin{figure}
\plotone{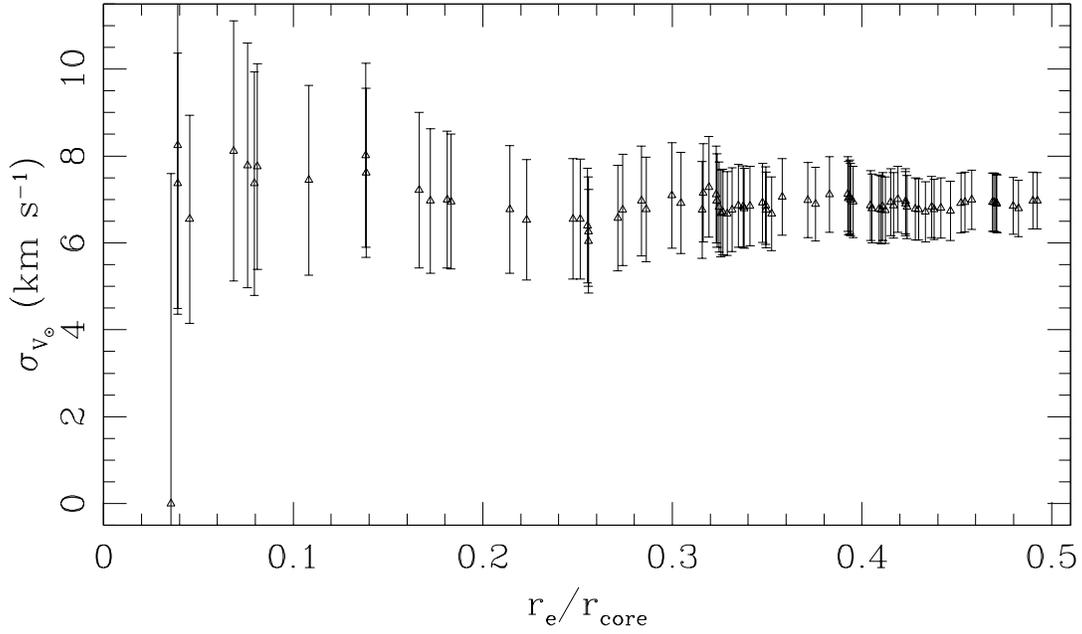}
\caption{Central velocity dispersion as a function of radius expressed
in terms of the King core radius. The dispersion is calculated so that
stars are added one by one to the central bin (except the first point which includes the 
innermost three stars in the dataset) and using the maximum likelihood method. A value
of $6.97\pm0.65$ km s$^{-1}$ is reached at about half the core radius. The dispersion
at this point includes the innermost 87 stars.}
\end{figure}

\begin{figure}
\plotone{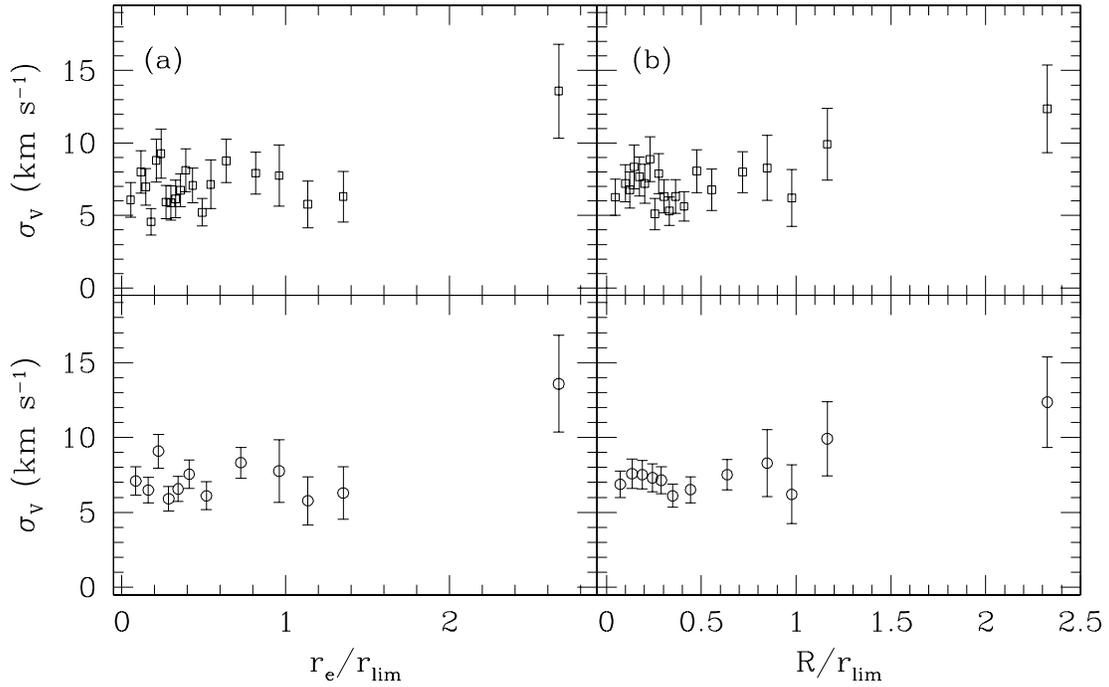}
\caption{Velocity dispersion versus ellitical distance (left panels) and
circular distance (right panels) for 23 and 46 stars per bin 
(lower
and upper panels). To take into account the fact that the outer regions have
a much lower density and therefore less stars are found there,
the last four dispersion points in each panel were calculated
with 10 stars each.}
\end{figure}

\begin{figure}
\plotone{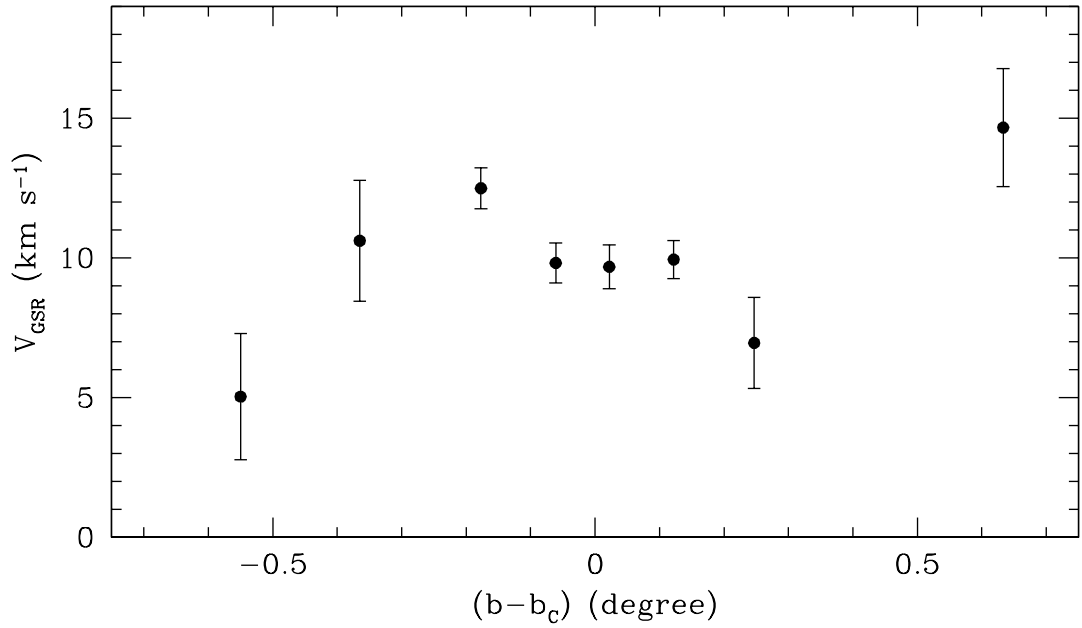}
\caption{Mean Radial Velocity trend (in Galactic Standard of Rest) for the Carina dSph.
The two outermost points on both sides from the center of Carina 
were calculated with 10 stars to take into account the lower density
in the outer regions of the dSph. 
A peak-to-peak difference of $\sim10$ km s$^{-1}$ is observed over a 1.2 degree scale (2.1 kpc).
We interpret this feature as indicative of tidal interaction.}
\end{figure}

\begin{figure}
\plotone{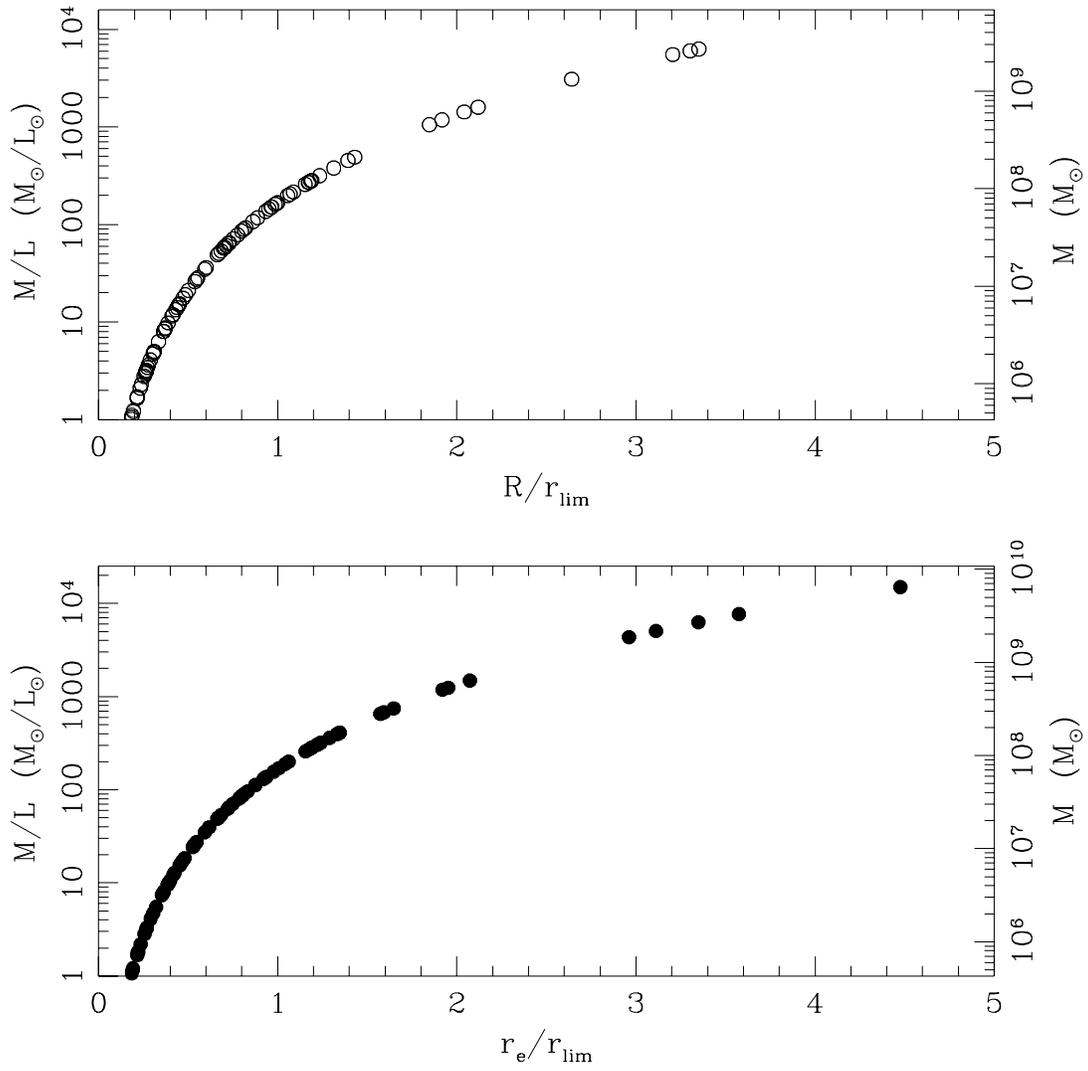}
\caption{The implied global $M/L$ implied for Carina as progressively more widely separated RV
members are attributed as bound to the satellite.  Open circles show the implied
$M/L$ of Carina assuming a spherical potential and the star's linear
projected distance from the center of Carina.  Solid circles
show the $M/L$ implied using  {\it elliptical radii} on the major axis.}
\end{figure}

\begin{figure}
\plotone{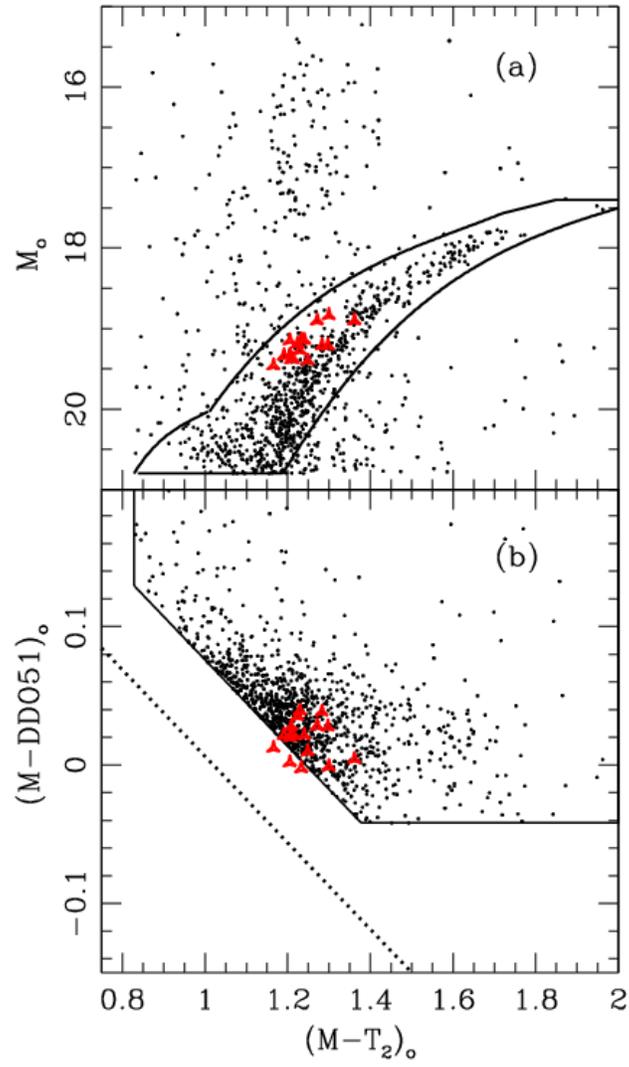}
\caption{(a) Color-magnitude diagram of all giant candidates selected by the method in Figure 3b.
The selection criterion applied to select Carina RGB stars from among all giant candidates is shown. 
Red triangles 
represent the 332 km s$^{-1}$ moving group stars. (b) Color-color diagram
corresponding to the CMD shown in (a). The \citeauthor*{Paper I} giant box is also shown as dotted lines.}
\end{figure}

\begin{figure}
\plotone{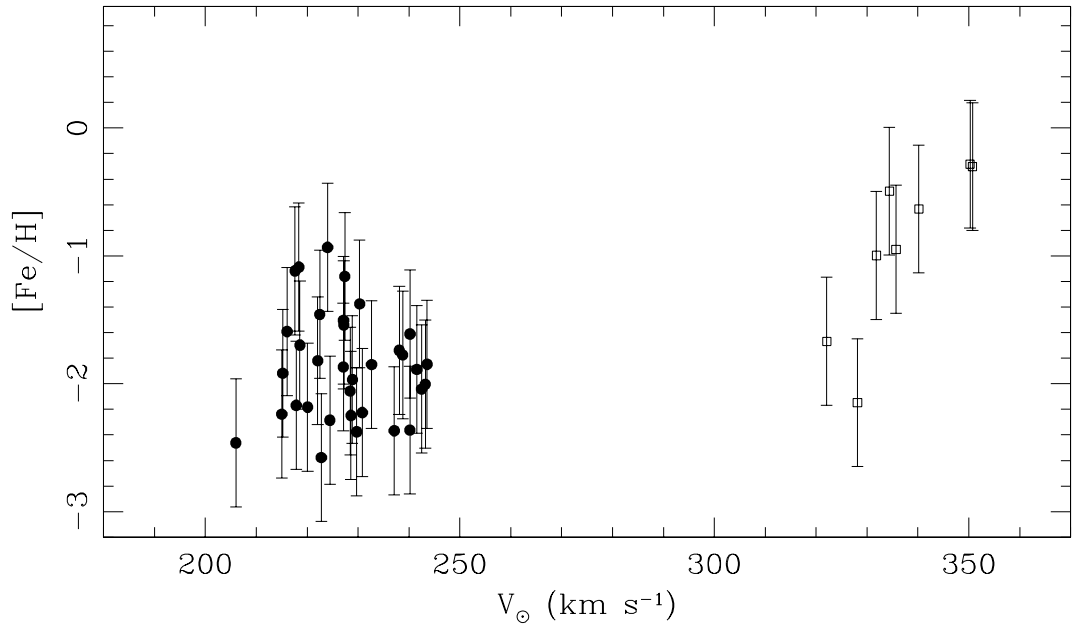}
\caption{$[Fe/H]$ from spectroscopic indices versus $V_{\sun}$.
The new moving group stars are clearly more metal rich than the 
Carina stars, indicating that they might belong to
a different population. The metallicities were calculated assuming
that all the stars in the Figure are at the distance of Carina.}

\end{figure}

\begin{figure}
\plotone{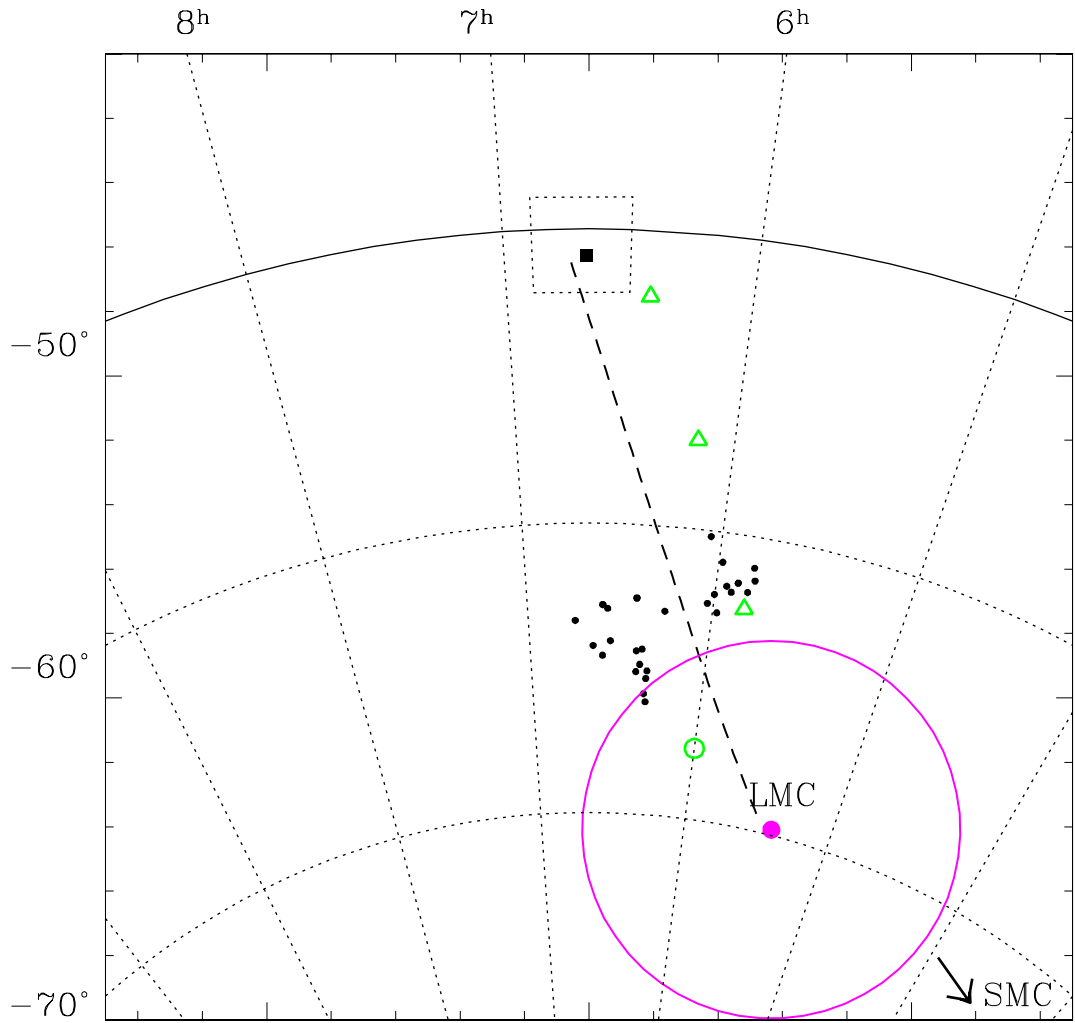}
\caption{Map of the fields in which we have found stars with velocities
like the LMC and the 332 km s$^{-1}$ group.  
The position of the LMC is marked with a magenta circle while the 
large circle around it
shows the 8$^{\circ}$ radius that marks the known extent of the 
disk.  The dotted quadrangle
around Carina (dSph center marked by the solid square symbol)
shows the approximate extent of the current photometric survey (Fig. 1).
Open green symbols show the location
of other fields studied as part of a separate study of the LMC 
halo (Nidever et al., in preparation), with the 
open green circle closest to the LMC marking the field used to make 
the CMD and 2CD shown in Figure 18.  The dashed line connecting the LMC 
with the Carina field marks a line of constant position angle 
between the two galaxies. 
The small black circles represent carbon stars in that region
found by \citet{kunkel1997a,kunkel1997b}.  The arrow points toward the 
center of the Small Magellanic Cloud.}
\end{figure}

\begin{figure}
\plotone{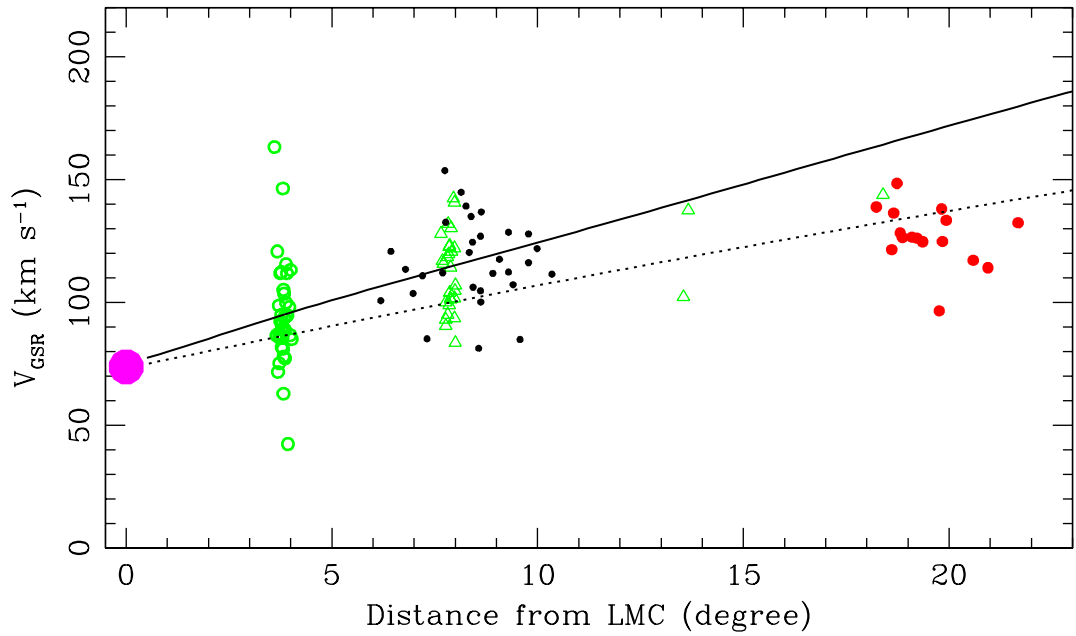}
\caption{Velocity trend for stars observed in the periphery of the
LMC and in the direction of Carina  as a
function of Galactic longitude and latitude. The LMC is indicated by the
big solid symbol at V$_{GSR}\sim75$. Red filled circles show
the 332 km s$^{-1}$ group observed with MIKE and GIRAFFE, open green triangles, open
green circles and black dots correspond to stars in the fields marked with the same 
symbols in Figure 16. The dashed/solid line shows the velocity trend 
expected for the halo/disk of the LMC extrapolated to the distance of 
Carina (\citeauthor*{vdM02}).}
\end{figure}

\begin{figure}
\plotone{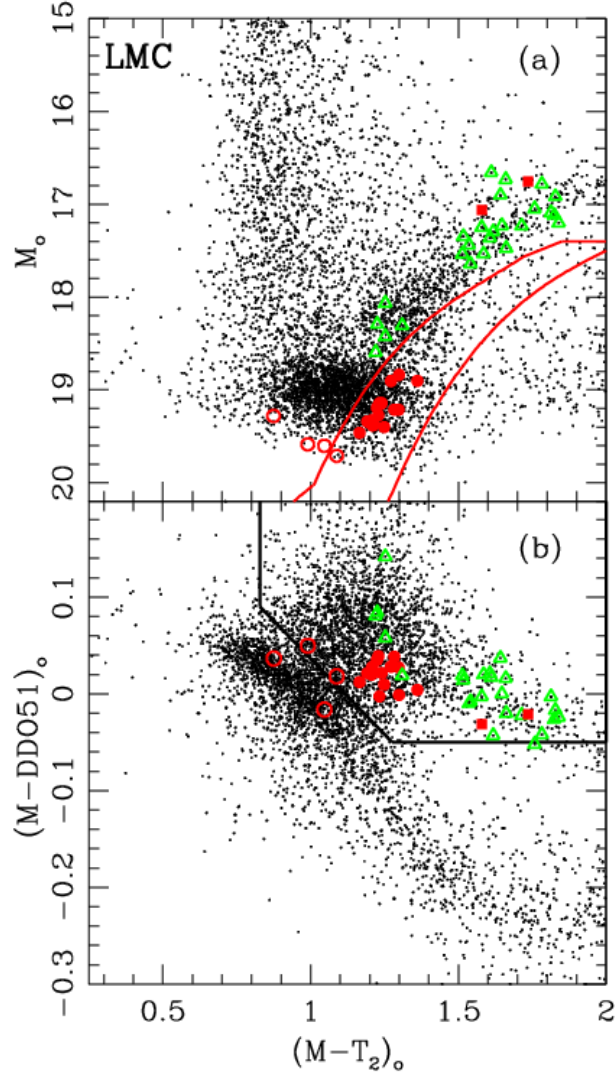}
\caption{Color-magnitude diagram and color-color diagram  for the field
in the periphery of the LMC shown as the green open circle in Figure 16. The 
red lines mark the RGB selection box of
Carina. The red circles represent members of 
the 332 km s$^{-1}$ group found in the Carina field, with solid 
circles marking those stars observed with
MIKE that satisfied our {\it Carina} selection criteria and the 
open red circles those VLT+GIRAFFE
stars with velocities consistent with membership in the 
332 km s$^{-1}$ group.  The green triangles
represent other Magellanic stars stars found in our separate LMC survey. 
The two red squares represent those two ``Magellanic giant candidates" 
observed with MIKE
in August 2005; both are found to have RVs consistent with 
membership in the 332 km s$^{-1}$ group.}
\end{figure}


\begin{thebibliography}{}
  
\bibitem[Armandroff \& Zinn(1988)Armandroff \& Zinn]{a88}
  Armandroff, T. E., \& Zinn, R. 1988, \aj, 96, 92


\bibitem[Armandroff et al.(1995)Armandroff, Olszewski \&
  Pryor]{Armandroff1995} Armandroff, T. E., Olszewski, E. W. \& Pryor,
  C. 1995, \aj, 110, 2131

\bibitem[Bailyn \& Steinmetz(2005)]{bailyn2005} Bailyn, J. \& Steinmetz, M. 2005, \apj, 627, 647


\bibitem[Binney \& Tremaine(1987)]{BT1987} Binney, J., \& Tremaine, S. 1987,
 in Galactic Dynamics, Princeton, New Jersey

\bibitem[Bullock \& Johnston(2005)]{BJ2005} Bullock, J. \& Johnston, K. V. 2005, \apj, 635, 931

\bibitem[Burkert(1997)]{Burkert1997} Burkert, A. 1997, \apj, 474,
  L99

\bibitem[Carlberg et al.(1997)]{Carlberg1997} Carlberg, R. G., Yee, H. K. C., \& Ellingson, 
E. 1997, \apj, 478, 462

\bibitem[Cenarro et al.(2001a)]{cenarro1} Cenarro, A. J., Cardiel, N., Gorgas, J., Peletier, R. F., Vazdekis, A. \& Prada, F.
2001, \mnras, 326, 959

\bibitem[Cenarro et al.(2001b)]{cenarro2} Cenarro, A. J., Gorgas, J., Cardiel, N., Pedraz, S., Peletier, R. F. \& Vazdekis, A.
2001, \mnras, 326, 981


\bibitem[Clewley et al.(2005)]{Clewley2005} Clewley, L., Warren, S. J.,
Hewett, P. C., Norris, J. E., Wilkinson, M. I. \& Evans, N. W. 2005, \mnras, 362, 349

\bibitem[Cole et al.(2004)]{cole04} Cole, A. A., Smecker-Hane, T. A., Tolstoy, E., Bosler, T. L. \&
Gallagher, J. S.  2004, \mnras, 347, 367
  
\bibitem[Cole et al.(2005)]{Cole05} Cole, A.~A., Tolstoy, E., 
Gallagher, J.~S., \& Smecker-Hane, T.~A. 2005, \aj, 129, 1465 
 
\bibitem[Choi et al.(2002)]{Choi2002} Choi. P. I., Guhathakurta, P. \&  
Johnston, K. V. 2002, \aj, 124, 310

\bibitem[Crane et al.(2003)]{Crane2003} Crane, J. D., Majewski, S. R., Rocha-Pinto, H. J.,
Frinchaboy, P. M., Skrutskie, M. F., \& Law, D. R. 2003, \apj, 594, L119

\bibitem[Cuddeford \& Miller(1990)]{Cuddeford1990} Cuddeford, P. \& Miller, J. C.
1990, \mnras, 244, 64

\bibitem[Dejonghe(1987)]{Dejonghe1987} Dejonghe, H. 1987, \mnras, 224, 13


\bibitem[Duffau et al.(2006)]{Duffau2006}
Duffau, S., Zinn, R., Vivas, A. K., Carraro, G., M{\'e}ndez, R. A., Winnick, R. 
\& Gallart, C. 2006, \apj, 636, L97

\bibitem[Faber \& Lin(1983)]{Faber1983} Faber, S. M. \& Lin, D. N. C. \apj,
266L, 17

 
\bibitem[Ferrara \& Tolstoy(2000)]{Ferrara2000} Ferrara, A., \& Tolstoy, E.\ 2000, \mnras, 313, 291

\bibitem[Fleck \& Kuhn(2003)]{fleck2003} Fleck, J. \& Kuhn, J. R. 2003,
  \apj, 592, 147

\bibitem[Frinchaboy et al.(2005)]{Frinchaboy2005} Frinchaboy, P. M., Mu\~{n}oz, R.
R., Majewski, S. R., Friel, E. D., Phelps, R. L.,
\& Kunkel, W. E.  2005, in ``Chemical Abundances and Mixing in Stars in
the Milky Way and its Satellites'',
eds. L. Pasquini \& S. Randich, ESO Astrophysics Symposia, {\it in press}
(astro-ph/0411127)


\bibitem[Frinchaboy et al.(2006)]{Frinchaboy2006} Frinchaboy, P. M., Mu\~{n}oz, R. R., Phelps, R. L.,
Majewski, S. R., \& Kunkel, W. E. 2006, \aj, 131, 922

\bibitem[Font et al.(2006)]{Font2006} Font, A., Johnston, K. V., Bullock, J. \& Robertson, B. 2006, \apj, 638, 585


\bibitem[Fusi Pecci et al.(1995)]{fusi95} Fusi Pecci, F., 
Bellazzini, M., Cacciari, C., \& Ferraro, F.~R.\ 1995, \aj, 110, 1664 

\bibitem[Geisler(1996)]{geisler96} Geisler, D. 1996, \aj, 111, 480G

\bibitem[Gilmore(2004)]{Gilmore2004} Gilmore, G. 2004, in
``Milky Way Surveys: The Structure and Evolution of our Galaxy'', ASP Conference
Proceedings Vol. 317, eds. D. Clemens, R. Shah, \& T. Brainerd (San Francisco: ASP), 239


\bibitem[G{\'o}mez-Flechoso et al.(1999)]{GFM99} G{\'o}mez-Flechoso, M. A.,
  Fux, R., \& Martinet, L. 1999, \aap, 347, 77

\bibitem[Hargreaves et al.(1994)]{Hargreaves1994} Hargreaves, J. C., Gilmore, G., 
Irwin, M. J., \& Carter, D. 1994, \mnras, 269, 957 


\bibitem[Hayashi et al.(2003)]{Hayashi2003} Hayashi, E., Navarro, J. F.,
  Taylor, J. E.,Stadel, J., \& Quinn, T. 2003, \apj, 584, 541

\bibitem[Hodge \& Michie(1969)]{hm69} Hodge, P. W. \& R. W. Michie 1969, \aj,
  74, 587

\bibitem[Ibata et al.(2001)]{Ibata2001} Ibata, R. A., Irwin, M. J.,
Lewis, G. F., Ferguson, A. M. N., \& Tanvir, N. 2001, Nature, 412, 49


\bibitem[Ibata et al.(2003)]{Ibata2003} Ibata, R. A., Irwin, M. J.,
Lewis, G. F., Ferguson, A. M. N., \& Tanvir, N. 2003, \mnras, 340, L21

\bibitem[Illingworth(1976)]{Illingworth76} Illingworth, G. 1976, \apj, 204, 73

\bibitem[Irwin \& Hatzidimitriou(1995)IH95]{IH95} Irwin, M. \&
  Hatzidimitriou, D. 1995, \mnras, 277, 1354 (IH95)


\bibitem[Johnston et al.(1999)]{JSH99}Johnston, K.  V.,
Sigurdsson, S. \& Hernquist, L. 1999, \mnras, 302, 771

\bibitem[Johnston et al.(2002)]{JCG02}Johnston, K.  V.,
Choi, P. I. \& Guhathakurta, P. 2002, \aj, 124, 127

\bibitem[Kalirai et al.(2005)]{Kalirai2005} Kalirai, J. S., Guhathakurta, P., Gilbert, K. M.,
Reitzel, D. B., Majewski, S. R., Rich, R. M., \& Cooper, M. C. 2005, (astro-ph/0512161)


\bibitem[Kauffmann et al.(1993)]{Kauffmann1993} Kauffmann, G., White, S.
  D. M., \& Guiderdoni, B. 1993, \mnras, 261, 201

\bibitem[Kazantzidis et al.(2004)]{Kazantzidis2004} Kazantzidis, S., Mayer, L., Mastropietro, C.,
 Diemand, J., Stadel, J., \& Moore, B. 2004, \apj, 608, 663


\bibitem[King(1966)]{King1966} King, I. R. 1966, \aj, 71, 64

\bibitem[King(1962)]{King1962} King, I. R. 1962, \aj, 136, 784

\bibitem[Kleyna et al.(1999)]{Kleyna1999} Kleyna, J. T., Geller, M. J.,
  Kenyon, S. J. \& Kurtz, M. J. 1999, \aj, 117,
  1275
  
\bibitem[Kleyna et al.(2002)]{Kleyna2002} Kleyna, J. T., Wilkinson, M.
  I., Evans, N. W., Gilmore, G. \& Frayn, C. 2002, \mnras, 330, 792

\bibitem[Kleyna et al.(2004)]{Kleyna2004} Kleyna, J. T., Wilkinson, M.
  I., Evans, N. W., \& Gilmore, G.  2004, \mnras, 354, L66

\bibitem[Klypin et al.(1999)]{Klypin1999} Klypin, A., Kravtsov, A. V.,
 Valenzuela, O., \& Prada, F. 1999, \apj, 522, 82

\bibitem[Kochanek(1996)]{kochanek96} Kochanek, C. S. 1996, \apj, 457, 228


\bibitem[Kroupa(1997)]{Kroupa1997} Kroupa, P. 1997, New Astronomy, 2, 139

\bibitem[Koch et al.(2006)]{Koch06} Koch, A., Grebel, E. K., Wyse, R. F. G., Kleyna, J. T.,
Wilkinson, M. I., Harbeck, D., Glimore, G. F. \& Evans, N. W.
2006, 131, 895

\bibitem[Kuhn \& Miller(1989)]{kuhn1989} Kuhn, J. R. \& Miller, R. H.
  1989, \apjl, 341, 41

\bibitem[Kuhn(1993)]{kuhn1993} Kuhn, J.~R.\ 1993, \apjl, 409, L13

\bibitem[Kuhn et al.(1996)Kuhn, Smith \& Hawley]{kuhn96} Kuhn, J. R., 
Smith, H. A. \& Hawley, S. L. 1996, \apjl, 469, L93 

\bibitem[Kunkel(1979)]{kunkel79} Kunkel, W. E. 1979, \apj, 228, 
718 
 
\bibitem[Kunkel et al.(1997a)]{kunkel1997a} Kunkel, W. E., Irwin, M. J. \& Demers, S. 1997a,
\aaps, 122, 463 

\bibitem[Kunkel et al.(1997b)]{kunkel1997b} Kunkel, W. E., Demers, S., Irwin, M. J.
\& Albert, L. 1997b, \apj, 488, L129


\bibitem[{\L}okas(2002)]{Lokas2002} {\L}okas, E. L., 2002, \mnras, 333, 697

\bibitem[{\L}okas et al.(2005)]{Lokas2005} {\L}okas, E. L., Mamon, G. A. \& Prada, F. 
2005, \mnras, 363, 918

\bibitem[Lynden-Bell(1982)]{lyndel82} Lynden-Bell, D.\ 1982, The 
Observatory, 102, 202 
 
\bibitem[Lynden-Bell \& Lynden-Bell(1995)]{LandL95} 
Lynden-Bell, D., \& Lynden-Bell, R.~M.\ 1995, \mnras, 275, 429 

\bibitem[Mac Low \& Ferrara(1999)]{MLF1999} Mac Low, M., \& Ferrara, A.\ 1999, \apj, 513, 142

\bibitem[Majewski(1994)]{maj94} Majewski, S.~R.\ 1994, \apjl, 431, L17 

\bibitem[Majewski et al.(1996)]{MPR96} Majewski, S. R., Phelps, R. \& Rich, R. M. 1996,
in The History of the Milky Way and Its Satellite System, ed. A. Burkert, D. Hartmann, \& S. Majewski,
(San Francisco: ASP), 1

\bibitem[Majewski et al.(2000a)Paper I]{Paper I} Majewski, S. R., Ostheimer, J.
  C., Kunkel, W. E. \& Patterson, R. J. 2000a, \aj, 120, 2550 (Paper I)

\bibitem[Majewski et al.(2000b)Paper II]{Paper II} Majewski, S. R., Ostheimer, J.
  C., Patterson, R. J., Kunkel, W. E., Johnston, K. V. \& Geisler, D.
  2000b, \aj, 119, 760 (Paper II)
  
\bibitem[Majewski et al.(2002)]{Majewski2002} Majewski, S. R., et al. 2002, in
``Modes of Star Formation and the Origin of Field Populations'', ASP Conference Proceedings Vol. 285, eds.
E. K. Grebel \& W. Brandner (San Francisco: ASP), 199

\bibitem[Majewski et al.(2003)]{MSWO} Majewski, S. R., Skrutskie, M.
  F., Weinberg, M. D. \& Ostheimer, J. C. 2003, \apj, 599, 1082

\bibitem[Majewski et al.(2004a)]{Majewski2004a} Majewski, S. R., et al. 2004a, \aj, 128, 245

\bibitem[Majewski(2004b)]{Majewski2004b} Majewski, S.~R.\ 2004b, 
Publications of the Astronomical Society of Australia, 21, 197 
 
\bibitem[Majewski et al.(2005)Paper VI]{Paper VI} Majewski, S. R., et al. 2005, \aj, 130, 2677
  
\bibitem[Mashchenko et al.(2005a)]{Mashchenko2005a} Mashchenko, S., 
Couchman, H. M. P. \& Sills, A. 2005a, \apj, 624, 726

\bibitem[Mashchenko et al.(2005b)]{Mashchenko2005b} Mashchenko, S., Sills, A., \& 
Couchman, H. M. P. 2005b, (astro-ph/0511567)

\bibitem[Mateo et al.(1993)]{Mateo1993} Mateo, M., Olszewski, E. W.,
Pryor, C., Welch, D. \& Fischer, P. 1993, \aj, 105, 510M

\bibitem[Mateo(1997)]{Mateo1997} Mateo, M. 1997, ASP
  Conf.~Ser.~116: The Nature of Elliptical Galaxies; 2nd Stromlo
  Symposium, 259


\bibitem[Mateo(1998)]{Mateo1998} Mateo, M. L. 1998, \araa, 36, 435 

\bibitem[Mayer et al.(2002)]{Mayer2002} Mayer, L., Moore, B., 
Quinn, T., Governato, F. \& Stadel, J. 2002, \mnras, 336, 119 

\bibitem[Milgrom(1995)]{Milgrom1995} Milgrom, M. 1995, \apj, 455, 439

\bibitem[Momany \& Zaggia(2005)]{momany2005} Momany, Y., \& Zaggia, S. 2005, \aap, 437, 339

\bibitem[Monelli et al.(2003)]{monelli2003} Monelli, M., et al. 2003, \aj, 126, 218

\bibitem[Monelli et al.(2004)]{monelli04} Monelli, M., et al. 2004,
Memorie della Societa Astronomica Italiana Supplement, 5, 65


\bibitem[Moore et al.(1999)]{Moore1999} Moore, B., Ghigna, S.,
  Governato, F., Lake, G., Quinn, T., Stadel, J., \& Tozzi, P. 1999,
  \apjl, 524, L19

\bibitem[Mu\~noz et al.(2005)]{Mu05} Mu\~noz, R. R. et al. 2005 \apjl, 631, L137 

\bibitem[Morrison et al.(2001)]{mor01} Morrison, H. L., Olszewski, E.
W., Mateo, M., Norris, J. E., Harding, P., Dohm-Palmer, R. C. \&
 Freeman, K. C. 2001, \aj, 121, 283

\bibitem[Newberg et al.(2002)]{Newberg2002} Newberg, H.  J. et
  al.\ 2002, \apj, 569, 245
  
\bibitem[Odenkirchen et al.(2001)]{Odenkirchen2001} Odenkirchen, M. et
  al.\ 2001, \aj, 122, 2538

\bibitem[Oh et al.(1992)]{Oh1992}  Oh, K.~S., Lin, D.~N.~C., \& Aarseth, S.~J.\ 1992, \apj, 386, 506


\bibitem[Oh et al.(1995)]{Oh95} Oh, K.~S., Lin, 
D.~N.~C., \& Aarseth, S.~J.\ 1995, \apj, 442, 142 
  
\bibitem[Olszewski et al.(1996)Olszewski, Pryor \&
  Armandroff]{Olszewski1996} Olszewski, E. W., Pryor, C. \& Armandroff,
  T. E. 1996, \aj, 111, 750
  
\bibitem[Olszewski \& Aaronson(1985)]{Olszewski1985} Olszewski, E. W. \&
  Aaronson, M. 1985, \aj, 90, 2221

\bibitem[Osterbrock et al.(1996)]{Osterbrock1996} Osterbrock, D.~E.,
Fulbright, J.~P., Martel, A.~R., Keane, M.~J., Trager, S.~C., \& Basri, G.\
1996, \pasp, 108, 277

\bibitem[Osterbrock et al.(1997)]{Osterbrock1997} Osterbrock, D.~E.,
Fulbright, J.~P., \& Bida, T.~A.\ 1997, \pasp, 109, 614


\bibitem[Palma et al.(2002)Palma, Majewski \& Johnston]{palma2002} Palma, 
C., Majewski, S. R. \& Johnston, K. V. 2002, \apj, 564, 736 

\bibitem[Piatek \& Pryor(1995)]{PP95} Piatek, S., \& Pryor, C.\ 1995, \aj, 109, 1071

\bibitem[Piatek et al.(2003)]{Piatek2003} Piatek, S., Pryor, C., Olszewski, E. W., 
Harris, H. C., Mateo, M., Minniti, D., \& Tinney, C. G.\ 2003, \aj, 126, 2346

\bibitem[Pryor \& Meylan(1993)]{Pryor1993} Pryor, C., \& Meylan, G.\ 1993, in ASP Conf. Ser. 50, 
Structure and Dynamics of Globular Clusters, ed. S. Djorgovski \& G. Meylan (San Francisco: ASP), 357

\bibitem[Pryor(1996)]{Pryor96} Pryor, C.\ 1996, ASP Conf.~Ser.~ 92:
Formation of the Galactic Halo...Inside and Out, 92, 424

\bibitem[Read et al.(2005a)]{Read2005a} Read, J. I., Wilkinson, M. I., Evans, N. W.,
Gilmore, G.  \& Kleyna, J. T. 2005a, (astro-ph/0506687)

\bibitem[Read et al.(2005b)]{Read2005b} Read, J. I., Wilkinson, M. I., Evans, N. W.,
Gilmore, G.  \& Kleyna, J. T. 2005b, (astro-ph/0505226)

\bibitem[Rocha-Pinto et al.(2003)]{RP2003} Rocha-Pinto, H. J., Majewski, S. R., Skrutskie, M. F., 
\& Crane, J. D. 2003, \apj, 594, L115 

\bibitem[Rocha-Pinto et al.(2004)]{RP2004} Rocha-Pinto, H. J., Majewski, S. R., Skrutskie, M. F., Crane, J. D.,
\& Patterson, R. J. 2004, \apj, 615, 732

  
\bibitem[Richstone \& Tremaine(1986)]{Richstone1986} Richstone, D. O. \&
  Tremaine, S. 1986, \aj, 92, 72

\bibitem[Salaris \& Girardi(2002)]{salaris2002} Salaris, M. \& Girardi, L. 2002, \mnras,
337, 332


\bibitem[Sanders \& McGaugh(2002)]{sanders2002} Sanders, R. H. \& McGaugh, S. S. 2002, \araa, 
40, 263

\bibitem[Schommer et al.(1992)]{schommer1992} Schommer, R. A., Suntzeff, N. B., Olszewski, E. W.,
\& Harris, H.C. 1992, \aj, 103, 447
 
\bibitem[Schlegel et al.(1998)]{Schlegel98} Schlegel, D. J., 
Finkbeiner, D. P., \& Davis, M. 1998, \apj, 500, 525

\bibitem[Smecker-Hane et al.(1996)]{S-H1996} Smecker-Hane, T. A., Stetson, P. B., Hesser, J. E.,
\& Vandenberg, D. A. 1996, in ``From Stars to Galaxies: the Impact of Stellar Physics
on Galaxy Evolution'', ASP Conference Proceedins Vol 98, eds. C. Leitherer, U. Fritze-von 
Alvensleben, \& J. Huchra (San Francisco: ASP), 328
 
\bibitem[Sohn et al.(2006)]{Sohn2006} Sohn, S., Majewski, S. R., Mu\~noz, R. R., Kunkel, W. E., 
Johnston, K. V., Ostheimer, J. C., Guhathakurta, P., Patterson, R. J.,
Siegel, M.H., \& Cooper, M. 2006, \apj, {\it submitted}

\bibitem[Stetson(1987)]{Stetson1987} Stetson, P. B. 1987, \pasp, 99, 191

\bibitem[Stoehr et al.(2002)]{Stoehr2002} Stoehr, F., White, S. D. M.,
  Tormen, G., \& Springel, V., 2002, \mnras, 335, L84


\bibitem[Tolstoy et al.(2003)]{Tolstoy2003} Tolstoy, E., Venn, K. A., Shetrone, M., 
Primas, F., Hill, V., Kaufer, A, \& Szeifert, T. 2003, \aj, 125, 707
  617, L119

\bibitem[Tolstoy et al.(2004)]{Tolstoy2004} Tolstoy el at., 2004, \apj,
  617, L119

\bibitem[Vader(1986)]{Vader1986} Vader., J. P. 1986, \apj, 305, 669  

\bibitem[Vader(1987)]{Vader1987} Vader., J. P. 1987, \apj, 317, 128 


\bibitem[Unavane et al.(1996)]{UWG1996} Unavane, M.,
Wyse, R. F. G., \& Gilmore, G. 1996, \mnras, 278, 727


\bibitem[van der Marel et al.(2002)vdM02]{vdM02} van der Marel, R. P., Alves, D. R., 
Hardy, E., \& Suntzeff, N. B. 2002, \aj, 124, 2639 (vdM02) 

\bibitem[Walcher et al.(2003)]{Walcher2003} Walcher, C. J., Fried, J.
  W., Burkert, A., \& Klessen, R. S. 2003, \aap, 406, 847

\bibitem[Walker et al.(2005)]{Walker2005} Walker, M. G., Mateo, M., Olszewski, E. W.,
Bernstein, R. A., Wang, X., \& Woodroofe, M.  2005, \aj, {\it in press} (astro-ph/0511465)

\bibitem[Westfall et al.(2006)]{Westfall2006} Westfall, K. B., Ostheimer,
  J. C., Frinchaboy, P. M., Patterson, R. J., Majewski, S. R., \&
  Kunkel, W. E. 2006, \aj, 131, 375

\bibitem[Wilkinson et al.(2004)]{W04} Wilkinson, M. I., Kleyna, J.
  T., Evans, N. W., Gilmore, G. F., Irwin, M. J., \& Grebel, E. K. 2004,
  \apj, 611, L21

\bibitem[Xiao et al.(2005)]{xiao2005} Xiao, W., Woodroofe, M., Walker, M.,
Mateo, M., \& Olszewski, E. 2005, \apj, 626, 145

\bibitem[Zaritsky \& Lin(1997)]{Zaritsky1997} Zaritsky, D., \& Lin, D. N. C. 1997,\aj, 114, 2545


\end{thebibliography}
\end{document}